\documentclass{article}
\usepackage[active]{srcltx}
\usepackage{amsmath}
\usepackage{bbm}
\usepackage{amsthm}
\usepackage{amsfonts}
\usepackage{amssymb}
\usepackage{amscd}
\usepackage{a4wide}
\usepackage{bm}
\usepackage{latexsym}
\usepackage{color}
\usepackage{dsfont}
\usepackage{indentfirst}
\usepackage[latin1]{inputenc}

\def\qed{\hfill $\blacksquare$}

\newtheorem{definition}{\underline{Definition}}[section]
\newtheorem{proposition}[definition]{Proposition}
\newtheorem{theorem}[definition]{Theorem}
\newtheorem{lema}[definition]{Lemma}

\theoremstyle{definition}

\numberwithin{equation}{section}

\begin{document}

\begin{titlepage}
\begin{center}
{\LARGE \bf On the zero-field orbital magnetic susceptibility of Bloch electrons in graphene-like solids:\\
Some rigorous results.}

\medskip

\today
\end{center}

\begin{center}
\small{Baptiste Savoie\footnote{Institute of Mathematics 'Simion Stoilow' of the Romanian Academy, P.O. Box 1--764, RO-014700 Bucharest, Romania; e-mail: baptiste.savoie@gmail.com .}}

\end{center}

\vspace{0.5cm}

\begin{abstract}
Starting with a nearest-neighbors tight-binding model, we rigorously investigate the bulk zero-field orbital susceptibility of a non-interacting Bloch electrons gas in graphene-like solids at fixed temperature and density of particles. In the zero-temperature limit and in the semiconducting situation, we derive a complete expression which holds for an arbitrary number of bands with possible degeneracies. In the particular case of a two-bands gapped model, all involved quantities are explicitly written down. Besides the formula that we obtain have the special feature to be suitable for numerical computations since it only involves the eigenvalues and associated eigenfunctions of the Bloch Hamiltonian, together with the derivatives (up to the second order) w.r.t. the quasi-momentum of the matrix-elements of the Bloch Hamiltonian. Finally we give a simple application for the two-bands gapped model by considering the case of a dispersion law which is linear w.r.t. the quasi-momentum in the gapless limit. Through this instance, the origin of the singularity, which expresses as a Dirac delta function of the Fermi energy, implied by the McClure's formula in monolayer graphene is discussed.
\end{abstract}

\tableofcontents

\vfill

\noindent
PACS-2010 number: 51.60.+a, 73.20.-r, 73.22.Pr, 75.20.Ck, 75.30.Cr.

\medskip

\noindent
MSC-2010 number: 81Q10, 82B10, 82D20, 82D37.

\medskip


\end{titlepage}
\section{Introduction and the main results.}

\subsection{Introduction.}

Isolated for the first time in 2004 by Novoselov \textit{et al.} \cite{No}, graphene is one of the most intriguing solid because of its exceptional electronic properties which, in great part, originate from its purely two-dimensional structure; see \cite{CGPNG} for a review of these properties. Regarded to be a semimetal since the pioneer works of Wallace in 1947 \cite{Wall} (he was the first one who investigated the band structure of graphene through a nearest-neighbors tight-binding model), graphene has the feature to exhibit a very high diamagnetic susceptibility. Note that this characteristic is shared by most semimetals, and to a lesser degree, by most narrow-gap intrinsic semiconductors.\\
\indent From a theoretical viewpoint, the first attempts to account for this large diamagnetic behavior go back at least to the 50's with the works of Smoluchowski \cite{Smo} and Hove \cite{Ho}. Even though these papers only were concerned with graphite (viewed as a piling up of graphene sheets) which shows the same peculiarity than graphene (graphite is known as one of the strongest diamagnetic solids), they are nevertheless interesting in regards to the method. The starting point in \cite{Smo,Ho} is the formula of Peierls \cite[Eq. (62c)]{Pei} obtained in the tight-binding approximation (it amounts to the same thing to consider a single-band model) for the zero-field orbital susceptibility of conduction electrons at non-zero temperature, together with the Wallace band structure \cite{Wall} (two bands touching at the Brillouin zone corners: the lower one completely filled, the upper completely empty at zero temperature). However in 1953, the Peierls formula threw back into doubt by Adams \cite{Adams}. This latter pointed out the existence of other contributions coming from the interband effects induced by the magnetic field; but apart from particular cases in which these contributions are far from being negligible, no general formulas were derived. Note that the quest to a complete formula (including the interband contributions) of the zero-field susceptibility of Bloch eletrons in solids generated an intense activity up to the middle of 70's, see e.g. \cite{MK} for a review. Note also that the first rigorous investigations in this direction came as late as 1990, see \cite{HeSj}. Recently, Briet \textit{et al.} \cite{BCS} have rigorously derived a complete formula for the zero-field susceptibility of Bloch electrons at fixed temperature and density for both metals and semiconductors/insulators. They have also given a rigorous justification to the Landau-Peierls approximation in low temperature and density limit.\\
\indent In 1956, McClure wrote down in the high temperature regime a formula for the zero-field orbital susceptibility of Bloch electrons in \textit{a two-dimensional honeycomb lattice} (modeling a single-layer graphite) taking into account the interband effects, see \cite[Eq. (3.15)]{McC}. To achieve that, he first calculated the energy levels induced by the magnetic field for the two-dimensional Wallace band structure in \cite{Wall}. This band structure yields a dispersion law which is, as a first approximation, linear w.r.t. the quasi-momentum near the band-touching points at the Brillouin zone corners. Subsequently McClure used these calculations to work out the Helmholtz free energy from which he derived the zero-field orbital susceptibility. He stressed that \textit{'all the conduction electrons diamagnetism of the two-dimensional model is due to band-to-band transitions'}.\\
\indent Following \cite{McC}, most papers were mainly concentrated either on 'real' graphite, or on compounds involving graphite-like structure, see e.g. \cite{McC3,SJM,Sadisa,Sa}. All the same we mention that when dealing with (infinite) single-layer graphite model, Safran \textit{et al.} \cite{Sadisa}, and later on \cite{Sa}, recovered the McClure formula in a regime of small values of the chemical potential (they do not tend to show some restrictions on the temperature regime). However the question of the behavior of the zero-field orbital susceptibility at zero temperature is not addressed. It should be point out that the method used in \cite{McC3,SJM,Sadisa,Sa} basically is different from the one implemented in \cite{McC}. Indeed their starting point is a given formula for the zero-field susceptibility of a 3-D Bloch electrons gas, in that case the one derived by Fukuyama \cite{Fu}. Afterwards they carry out some transformations in that formula to take account of the dispersion law induced by the band structure of materials.\\
\indent In 1987, Levintovitch \textit{at al.} \cite{LeKo1} were interested in the zero-field orbital susceptibility of a \textit{two-dimensional Bloch electrons gas in narrow-gap semiconductors} at zero temperature and fixed density. Their method consisted in expressing a (quite) modified version of the zero-field susceptibility formula derived by Wannier \textit{et al.} \cite{WU} only in terms of a dispersion law corresponding to a two-bands gapped model (the origin of energy is taken at the middle of the gap, and when the gap width is zero, the two-bands model used by McClure in \cite{McC} is recovered). In the zero-temperature limit together with the semiconducting situation, they found that the zero-field orbital susceptibility solely is determined by the interband contributions, and further, it is in inverse proportion to the gap width, see \cite[Eqs. (22)-(23)]{LeKo1}. Besides in the gapless limit, it has a singularity expressed as a Dirac delta-like dependence on the Fermi energy (that is the zero-field orbital susceptibility is minus infinity at band-touching point, and vanishes otherwise). They claimed that their result is in good agreement with the McClure's formula in the zero-temperature limit.\\
\indent After the experimental discovery of Novoselov \textit{et al.} \cite{No}, many investigations arose to figure out the large diamagnetic behavior of monolayer graphene. Their approach substantially is different from the previous authors: it is based on the observation that the dispersion law (derived from the 2-D Wallace band structure in \cite{Wall}) in the neighborhood of the band-touching points in the Brillouin zone is, due to its linearity, analogous to the energy of massless relativistic particles. Since that suggests a massless Dirac fermions-like behavior of quasi-particles around the band-touching points, most papers were focused on the diamagnetism of massless Dirac fermions, see e.g. \cite{KA, Fu2, Nak, KA2}. We do not give further details about their method since this present paper will not deal with massless fermions (we anyway mention that the starting point in \cite{Fu2,Nak} is the zero-field  susceptibility formula derived by Fukuyama in \cite{Fu}). However we emphasize that for a gapless Dirac fermions system, Koshino \textit{et al.} and Fukuyama obtained that the zero-field orbital susceptibility at zero temperature is expressed as a delta function of the Fermi energy (the origin of energy is taken at band-touching point), see \cite[Eq. (37)]{KA} and \cite[Eq. (3c)]{Fu2}. Besides they stated that this result is in accordance with the McClure's results at finite temperature. In \cite{Nak}, Nakamura was interested in gapped Dirac fermions systems. In the semiconducting situation, he found that the zero-field orbital susceptibility at zero temperature is in inverse proportion to the energy gap, see
\cite[Eq. (7)]{Nak}. In the zero-energy gap limit, the results of \cite{KA, Fu2} are recovered. For the sake of completeness, we mention that within the framework of 2-D systems with a linear, massless Dirac-like dispersion law, the behavior of the zero-field susceptibility at zero-temperature (brought up just above) for such systems with and without a band gap, was already pointed out by Sharapov \textit{et al.} in 2004, see
\cite[Eqs. (7.4)-(7.6)]{SGB2} and also \cite{SGB1}.\\
\indent In the light of the above mentioned results, the singular behavior of the zero-field orbital susceptibility at zero temperature of Bloch electrons in graphene seems to be regularized when opening a gap in the band structure of graphene. This brings us naturally to be interested in the zero-field orbital susceptibility of gapped graphene to shed light on the possible origins of the delta-like singularity, other than the purely semimetallic nature of graphene at zero temperature. Furthermore, the study of fundamental properties of gapped graphene has recently found a renewal of interest due to new technical processes allowing to open a band gap in graphene materials, see e.g. \cite{2P, 2P2}.\\

\indent In this present paper, we rigorously investigate the zero-field orbital susceptibility of a non-interacting Bloch electrons gas in graphene-like solids. By a graphene-like solid, we mean a purely two-dimensional crystal which is an intrinsic semiconductor (possibly with a narrow gap) at zero temperature. In the one-body approximation, our starting point is a nearest-neighbors tight-binding model for the Hamiltonian  of a Bloch electron in a finite-size graphene-like solid plunged into a constant magnetic field applied perpendicular to the solid plane. Our model of tight-binding Hamiltonian is a Harper-like operator (see e.g. \cite{N2,C1,CR1}), and is similar to the one of Brynildsen \textit{et al.} \cite{BrC} who lately have rigorously investigated the Faraday rotation in graphene-like solids. From the thermodynamic limit of the grand-canonical pressure (the question of its existence is there fully addressed, see Lemma \ref{thm1} below), we derive a general formula for the bulk zero-field orbital susceptibility at fixed positive temperature and density of electrons, see \eqref{suscepti4}. This is made possible by the use of the so-called gauge invariant magnetic perturbation theory (see e.g. \cite{C0,N1} and \cite{BrCoLo1, CNP, CN3, BS, BrC} for further applications), followed by the Bloch-Floquet decomposition. After carrying out some convenient transformations needed to perform the zero-temperature limit, then in the semiconducting situation, we get from \eqref{suscepti4} a complete formula for the zero-field orbital susceptibility at zero temperature and fixed density which holds for an arbitrary number of bands with possible degeneracies, see \eqref{xSC1}. It consists merely of \eqref{xSC2} when considering the particular case of a two-bands gapped model. The whole of terms involved in \eqref{xSC2} are explicitly written down and are expressed only in terms of eigenvalues and associated eigenfunctions of the Bloch Hamiltonian, together with the derivatives (up to the second order) w.r.t. the quasi-momentum of its matrix-elements. The terms involved in \eqref{xSC1} have the same special feature what makes our formulas suitable for numerical computations. Finally, we give an application for the two-bands gapped model by treating the case of a dispersion law which is linear w.r.t. the quasi-momentum in the gapless limit (this is a simple 'textbook model' for monolayer graphene). The singular behavior of the bulk zero-field orbital susceptibility at zero temperature obtained in the gapless limit is construed from a mathematical viewpoint.

\subsection{The setting and the main results.}

Consider a two-dimensional finite-size crystal subjected to a uniform external magnetic field applied orthogonal to the solid. By crystal, we mean an ideal crystalline ordered solid composed of a regular lattice of identical atoms whose nuclei are fixed at their equilibrium positions. Due to the magnetic field, the Bloch electrons possess an orbital magnetic moment together with a spin magnetic moment. Since we only are interested in orbital diamagnetic effects, we disregard the electron spin. Furthermore we neglect the self-interactions: the Bloch electrons gas is supposed strongly diluted. Besides the finite-size crystal is at equilibrium with a thermal and particles bath.\\
\indent Let us precise our assumptions. Let $\boldsymbol{\Upsilon} := \{\boldsymbol{\upsilon}:\boldsymbol{\upsilon} = \sum_{k=1}^{2} n_{k} \bold{a}_{k}; n_{k} \in \mathbb{Z}\}$ be a Bravais lattice generated by two independent $\mathbb{R}^{2}$-vectors $\bold{a}_{k}$, $k=1,2$. Let $\Omega := \{\bold{x} : \vert \bold{x} \vert \leq \inf_{\boldsymbol{\upsilon} \in \boldsymbol{\Upsilon} \setminus \{\bold{0}\}} \vert \bold{x} - \boldsymbol{\upsilon}\vert \}$ be the Wigner-Seitz (W-S) cell of $\boldsymbol{\Upsilon}$, and denote by $\Theta$ the finite set of vectors modeling the basis (i.e. the positions of the ions at equilibrium in $\Omega$). Hereafter, we suppose that $\mathrm{Card}(\Theta) \geq 2$. The finite-size ion-core mesh is given by $\Lambda_{N} := \{\underline{\bold{x}} + \boldsymbol{\upsilon}: \underline{\bold{x}} \in \Theta, \boldsymbol{\upsilon} = \sum_{k=1}^{2} n_{k}\bold{a}_{k}; \vert n_{k} \vert \leq N\}$, $N \in \mathbb{N}^{*}$. Note that it basically boils down to the same thing to consider a central region of $\Lambda :=  \Theta + \boldsymbol{\Upsilon} = \{\underline{\bold{x}}+ \boldsymbol{\upsilon}:  \underline{\bold{x}} \in \Theta,\boldsymbol{\upsilon} \in \boldsymbol{\Upsilon} \}$ which stands for the ion-core mesh of the corresponding infinite crystal. The magnetic field is given by the 3-dimensional vector $\bold{B}:=(0,0,B)$, $B \geq 0$. By seeing a $\mathbb{R}^{2}$-vector $\bold{x}=(x_{1},x_{2})$ as the 3-dimensional vector $(x_{1},x_{2},0)$, the magnetic potential vector which reigns in the crystal is defined by $B \bold{a}(\bold{x}) := \frac{1}{2} (0,0,B) \wedge (x_{1},x_{2},0)= \frac{B}{2}(-x_{2},x_{1},0)$ (we use the symmetric transverse gauge). Throughout this paper, we denote by $b:= eB/c \in \mathbb{R}^{+}$ the electron cyclotron frequency where we set $m_{\mathrm{electron}} = 1$. Note that below we set $\hbar=1$ too.\\
\indent Under the above conditions, let us introduce the one-particle Hamiltonian. To fix notation and give some additional assumptions, we firstly start with the situation in which the crystal is infinite.
In view of our discrete approach, let $l^{2}(\Lambda)$ be the one-particle Hilbert space and denote by $\{\delta_{\bold{x}}\}_{\bold{x} \in \Lambda}$ the canonical basis, where $\delta_{\bold{x}}(\bold{y}) = 1$ when $\bold{y}=\bold{x}$ and $0$ otherwise. Recall that the kernel of any bounded operators $A$ on $l^{2}(\Lambda)$ is given by $A(\bold{x},\bold{y}) := \langle \delta_{\bold{x}},A \delta_{\bold{y}}\rangle$ $(\bold{x},\bold{y}) \in \Lambda^{2}$, where $\langle\cdot\,,\cdot\,\rangle$ denotes the usual scalar product on $l^{2}(\Lambda)$.
In the one-body approximation, let $H_{0}$ be the Hamiltonian on $l^{2}(\Lambda)$ which determines the dynamics of each Bloch electron in the absence of the magnetic field. We denote by $H_{0}(\cdot\,,\cdot\,):\Lambda\times \Lambda \rightarrow \mathbb{C}$ the kernel of $H_{0}$. In the whole of this paper, we assume that $H_{0}$ satisfies the following three properties (given through its kernel):
\begin{itemize}
\item[(P1)] $H_{0}$ is a self-adjoint operator, i.e.:
\begin{equation*}
\forall (\bold{x},\bold{y}) \in \Lambda^{2},\quad H_{0}(\bold{x},\bold{y}) = \overline{H_{0}(\bold{y},\bold{x})}.
\end{equation*}
\item[(P2)] $H_{0}$ is a nearest-neighbour tight-binding operator, that is:
\begin{equation}
\label{nerneig}
\exists C>0\quad : \quad H_{0}(\bold{x},\bold{y}) = 0 \quad \textrm{when $\vert \bold{x} - \bold{y}\vert \geq C$}.
\end{equation}
\item[(P3)] $H_{0}$ commutes with the translations of the $\boldsymbol{\Upsilon}$ Bravais lattice, i.e.:
\begin{equation*}
\forall \boldsymbol{\upsilon} \in \boldsymbol{\Upsilon},\quad H_{0}(\bold{x}+\boldsymbol{\upsilon},\bold{y}+\boldsymbol{\upsilon}) = H_{0}(\bold{x},\bold{y}) \quad (\bold{x},\bold{y}) \in \Lambda^{2}.
\end{equation*}
\end{itemize}
Usually the magnitude of the non-zero hopping matrix-elements for $H_{0}(\cdot\,,\cdot\,)$ is fixed to some finite constant (typically, fixed to $1$). In this case, and we suppose that hereafter, the kernel $H_{0}(\cdot\,,\cdot\,)$ is \textit{EAD} (=Exponentially Almost Diagonal), i.e. there exist two constants $c_{1},c_{2}>0$ s.t.
\begin{equation}
\label{esH_0}
\forall (\bold{x},\bold{y}) \in \Lambda^{2},\quad \vert H_{0}(\bold{x},\bold{y})\vert \leq c_{1} \mathrm{e}^{-c_{2}\vert \bold{x} - \bold{y}\vert}.
\end{equation}
In presence of magnetic fields, the Hamiltonian $H_{b}$ on $l^{2}(\Lambda)$ is defined through its kernel $H_{b}(\cdot\,,\cdot\,) : \Lambda\times\Lambda \rightarrow \mathbb{C}$ which is obtained from $H_{0}(\cdot\,,\cdot\,)$ by using the so-called 'Peierls substitution':
\begin{equation}
\label{magnH_b}
\forall(\bold{x},\bold{y}) \in \Lambda^{2},\quad H_{b}(\bold{x},\bold{y}) := \mathrm{e}^{i b \phi(\bold{x},\bold{y})} H_{0}(\bold{x},\bold{y}) \quad b \in \mathbb{R}^{+},\, \hbar=1;
\end{equation}
here $\phi(\cdot\,,\cdot\,)$ stands for the usual magnetic phase defined by $\phi(\bold{x},\bold{y}) := \bold{x}\cdot \bold{a}(\bold{y}) = \frac{1}{2}(y_{1}x_{2} - x_{1}y_{2})$. Since the phase is antisymmetric, then $H_{b}$ is a self-adjoint operator by virtue of (P1). Moreover \eqref{nerneig} still holds true for $H_{b}(\cdot\,,\cdot\,)$, and due to (P3), $H_{b}$ commutes with the magnetic translations:
\begin{equation}
\label{transmag}
\forall \boldsymbol{\upsilon} \in \boldsymbol{\Upsilon},\quad H_{b}(\bold{x} + \boldsymbol{\upsilon},\bold{y} + \boldsymbol{\upsilon}) \mathrm{e}^{i b \phi(\bold{y},\boldsymbol{\upsilon})} = \mathrm{e}^{i b \phi(\bold{x},\boldsymbol{\upsilon})} H_{b}(\bold{x},\bold{y})\quad (\bold{x},\bold{y}) \in \Lambda^{2}.
\end{equation}
Defined via \eqref{magnH_b}, $H_{b}$ actually is a Harper-like operator; see e.g. \cite{N2,C1} for its spectral properties.\\
\indent Let us get back to our initial problem dealing with the Bloch electrons gas in the finite-size crystal. The one-particle Hamiltonian in presence of magnetic fields is defined from $H_{b}$ by:
\begin{equation}
\label{Hvfini}
H_{N,b} := \chi_{N} H_{b} \chi_{N} \quad b \in \mathbb{R}^{+},
\end{equation}
where $\chi_{N}$ denotes the characteristic function of the finite-size ion-core mesh $\Lambda_{N}$. For any $b \in \mathbb{R}^{+}$, \eqref{Hvfini} defines a family of bounded and self-adjoint operators on $l^{2}(\Lambda)$ for any $N \in \mathbb{N}^{*}$. Obviously this definition corresponds to choose Dirichlet boundary conditions on the edges of $\Lambda_{N}$. Hereafter we denote by $\{e_{b}^{(j)}\}_{j=1}^{N}$ the set of eigenvalues of $H_{N,b}$ counting multiplicities and in increasing order.\\
\indent Now define the pressure of the Bloch electrons gas in the finite-size crystal under the grand-canonical conditions. Recall that in the grand-canonical ensemble, the fixed external parameters are $(\beta,z,\mathrm{Card}(\Lambda_{N}))$, where: $\beta := (k_{B}T)^{-1} > 0$ is the 'inverse' temperature ($k_{B}$ stands for the Boltzmann constant) and $z:= \mathrm{e}^{\beta \mu} >0$ is the fugacity ($\mu \in \mathbb{R}$ stands for the chemical potential). For any $\beta>0$, $z>0$ and $b \geq 0$, the grand-canonical pressure is given by (see e.g. \cite{Ro}):
\begin{equation}
\label{pressurevf}
P_{N}(\beta,z,b) := \frac{1}{\beta} \frac{1}{\mathrm{Card}(\Lambda_{N})} \sum_{j=1}^{N} \ln ( 1 + z\mathrm{e}^{- \beta e_{b}^{(j)}}).
\end{equation}
\indent Let us introduce some additional notation. Let $R_{b}(\cdot\,,\cdot\,;\xi) : \Lambda \times \Lambda \rightarrow \mathbb{C}$ be the kernel of the resolvent operator $R_{b}(\xi):=(H_{b} - \xi)^{-1}$ $\forall \xi \in \rho(H_{b})$. Let $\Gamma$ be a simple positively oriented closed contour around the spectrum of $H_{b}$ defined as in \eqref{Gamma}. We set $\mathfrak{f}(\beta,z;\cdot\,) := \ln(1 + z \mathrm{e}^{-\beta \cdot\,})$ for any $\beta,z>0$.\\
\indent Our main technical result is given in the following lemma:
\begin{lema}
\label{thm1}
$\mathrm{(i)}$. For any $\beta >0$, $z >0$ and $b \geq 0$, the thermodynamic limit of the grand-canonical pressure exists. More precisely:
\begin{equation}
\label{limP}
P(\beta,z,b) := \lim_{N \rightarrow \infty} P_{N}(\beta,z,b) = \frac{1}{\beta} \frac{1}{\mathrm{Card}(\Theta)} \frac{i}{2\pi} \int_{\Gamma} \mathrm{d}\xi\, \mathfrak{f}(\beta,z;\xi) \sum_{\underline{\bold{x}} \in \Theta} R_{b}(\underline{\bold{x}},\underline{\bold{x}};\xi),
\end{equation}
uniformly in $(\beta,z) \in [\beta_{1},\beta_{2}] \times [z_{1},z_{2}]$, $0< \beta_{1}<\beta_{2}<\infty$ and $0<z_{1}<z_{2}<\infty$.\\
$\mathrm{(ii)}$. For any $\beta>0$ and $b \geq 0$, $P(\beta,\cdot\,,b)$ is a $\mathcal{C}^{\infty}$-function on $\mathbb{R}^{*}_{+}$.\\
$\mathrm{(iii)}$. For any $\beta >0$ and $z>0$, $P(\beta,z,\cdot\,)$ is a $\mathcal{C}^{2}$-function on $\mathbb{R}_{+}$.
\end{lema}

By virtue of Lemma \ref{thm1}, introduce the bulk density of particles defined by:
\begin{equation}
\label{rhobul}
\rho(\beta,z,b) := \beta z \frac{\partial P}{\partial z}(\beta,z,b), \quad \beta>0,\,z>0,\, b\geq 0,
\end{equation}
as well as the bulk orbital susceptibility defined as the second derivative of the bulk pressure w.r.t. the intensity $B$ of the magnetic field (see e.g. \cite{ABN,BrCoLo1,BCS,BS} and references therein):
\begin{equation}
\label{defcalX}
\mathcal{X}(\beta,z,b) := \bigg(\frac{e}{c}\bigg)^{2} \frac{\partial^{2} P}{\partial b^{2}}(\beta,z,b),\quad \beta>0,\,z >0,\, b\geq 0.
\end{equation}
From now on, the density of particles $\rho_{0}>0$ becomes our fixed external parameter. Seeing the bulk density $\rho$ as a function of the $\mu$-variable (instead of the $z$-variable), we denote by $\mu^{(0)}(\beta,\rho_{0},b) \in \mathbb{R}$ the unique solution of the equation $\rho(\beta,\mathrm{e}^{\beta \mu}, b) = \rho_{0}$. Note that the inversion of the relation between the bulk density and the chemical potential is ensured by the fact that $\mu \mapsto \rho(\beta,\mathrm{e}^{\beta \mu},b)$ is a strictly increasing function on $(0,\infty)$. Under these conditions, the bulk zero-field orbital susceptibility at fixed positive 'temperature' $\beta>0$ and density $\rho_{0}>0$ is defined by:
\begin{equation*}
\mathcal{X}(\beta,\rho_{0}) = \mathcal{X}(\beta,\rho_{0},0):= \mathcal{X}(\beta, \mathrm{e}^{\beta \mu^{(0)}(\beta,\rho_{0},b=0)},b=0).
\end{equation*}
\indent Before formulating the main results of this paper, let us introduce some additional notation.\\
Let $\boldsymbol{\Upsilon}^{*}$ be the dual lattice of $\boldsymbol{\Upsilon}$ generated by the vectors $\bold{b}_{l}$, $l=1,2$ defined from the $\bold{a}_{k}$'s by $\bold{a}_{k} \cdot \bold{b}_{l}= 2\pi \delta_{k,l}$. Denote by $\Omega^{*}$ the Brillouin zone of $\boldsymbol{\Upsilon}^{*}$. By virtue of (P3), the Floquet theory for periodic operators allows us to use the bands structure of the spectrum of $H_{0}$, see e.g. \cite{Ku}. Under the Bloch-Floquet unitary transformation, $l^{2}(\Lambda) \cong \int_{\Omega^{*}}^{\oplus} \mathrm{d}\bold{k}\, l^{2}(\Theta)$ and $H_{0}$ can be seen as the direct integral $\int_{\Omega^{*}}^{\oplus} \mathrm{d}\bold{k}\, H(\bold{k})$, see e.g. \cite[Sect. XIII.16]{RS4}. For each $\bold{k} \in \Omega^{*}$, the Bloch Hamiltonian $H(\bold{k})$ lives on $l^{2}(\Theta)$ and it is defined via its kernel which reads as (see e.g. \cite[Eq. (2.4)]{BCZ}):
\begin{equation*}
\label{Hokk1}
\forall(\underline{\bold{x}},\underline{\bold{y}}) \in \Theta^{2},\quad H(\underline{\bold{x}},\underline{\bold{y}};\bold{k}) := \sum_{\boldsymbol{\upsilon} \in \boldsymbol{\Upsilon}} \mathrm{e}^{-i \bold{k} \cdot (\underline{\bold{x}} + \boldsymbol{\upsilon} - \underline{\bold{y}})} H_{0}(\underline{\bold{x}} + \boldsymbol{\upsilon}, \underline{\bold{y}}) \quad \bold{k} \in \Omega^{*}.
\end{equation*}
Note that in the canonical basis of $l^{2}(\Theta)$, the matrix of $H(\bold{k})$ is nothing but $[H(\underline{\bold{x}}_{l},\underline{\bold{x}}_{m};\bold{k})]_{l,m}$, $1\leq l,m\leq \mathrm{Card}(\Theta)$. Introduce as well on $l^{2}(\Theta)$ the operators $(\partial_{k_{\alpha}} H)(\bold{k})$ and $(\partial_{k_{\gamma}} \partial_{k_{\alpha}} H)(\bold{k})$ with $\alpha,\gamma=1,2$ and $\bold{k} \in \Omega^{*}$, generated via their kernels respectively defined on $\Theta^{2}$ by:
\begin{gather}
\label{partialH}
(\partial_{k_{\alpha}} H)(\underline{\bold{x}},\underline{\bold{y}};\bold{k}) := \frac{\partial}{\partial k_{\alpha}} H(\underline{\bold{x}},\underline{\bold{y}};\bold{k}) =  - i \sum_{\boldsymbol{\upsilon} \in \boldsymbol{\Upsilon}} (\underline{x}_{\alpha} + \upsilon_{\alpha} - \underline{y}_{\alpha}) \mathrm{e}^{- i\bold{k} \cdot (\underline{\bold{x}} + \boldsymbol{\upsilon} - \underline{\bold{y}})} H_{0}(\underline{\bold{x}} + \boldsymbol{\upsilon},\underline{\bold{y}}),\\
\begin{split}
&(\partial_{k_{\gamma}} \partial_{k_{\alpha}} H)(\underline{\bold{x}},\underline{\bold{y}};\bold{k}) := \frac{\partial^{2}}{\partial k_{\gamma} \partial k_{\alpha}} H(\underline{\bold{x}},\underline{\bold{y}};\bold{k})= \\
&=  - \sum_{\boldsymbol{\upsilon} \in \boldsymbol{\Upsilon}} (\underline{x}_{\gamma} + \upsilon_{\gamma} - \underline{y}_{\gamma}) (\underline{x}_{\alpha} + \upsilon_{\alpha} - \underline{y}_{\alpha}) \mathrm{e}^{- i\bold{k} \cdot (\underline{\bold{x}} + \boldsymbol{\upsilon} - \underline{\bold{y}})} H_{0}(\underline{\bold{x}} + \boldsymbol{\upsilon},\underline{\bold{y}}) = \frac{\partial^{2}}{\partial k_{\alpha} \partial k_{\gamma}} H(\underline{\bold{x}},\underline{\bold{y}};\bold{k})\nonumber.
\end{split}
\end{gather}
Since $H_{0}(\cdot\,,\cdot\,)$ is \textit{EAD}, the kernels $(\partial_{k_{\alpha}} H)(\cdot\,,\cdot\,;\bold{k})$ and $(\partial_{k_{\gamma}} \partial_{k_{\alpha}} H)(\cdot\,,\cdot\,;\bold{k})$ are well-defined on $\Theta^{2}$. Getting back to the Bloch Hamiltonian, we denote by $\{E_{j}(\bold{k})\}_{j=1}^{\mathrm{Card}(\Theta)}$ the set of its eigenvalues, counting multiplicities and \textit{in increasing order}. Due to this choice of labelling, the Bloch energies $E_{j}(\cdot\,)$ are $\boldsymbol{\Upsilon}^{*}$-periodic and continuous functions but they are not differentiable at (possible) crossing-points; for a review see e.g.
\cite[Sect. III]{N}. Hereafter we will consider both situations:
\begin{itemize}
\item[(A1)] \label{x} The $E_{j}(\cdot\,)$'s, $j=1,\ldots,\mathrm{Card}(\Theta)$ are non-degenerate for Lebesgue-almost all $\bold{k} \in \Omega^{*}$.
\item[(A2)] \label{y} The $E_{j}(\cdot\,)$'s, $j=1,\ldots,\mathrm{Card}(\Theta)$ are non-degenerate everywhere on $\Omega^{*}$.
\end{itemize}
In other words, (A1) means that the set formed by the crossing-points has a zero Lebesgue-measure. As for (A2), it amounts to the same thing to assume that the Bloch bands are simple; the $j$-th Bloch band is defined by $\mathcal{E}_{j} := \bigcup_{\bold{k} \in \Omega^{*}} E_{j}(\bold{k})$.
Under either assumption, the spectrum of $H_{0}$ is absolutely continuous and given as a set of points by $\sigma(H_{0}) = \bigcup_{j=1}^{\mathrm{Card}(\Theta)} \mathcal{E}_{j}$. The energy bands of $H_{0}$ corresponds to the disjoint union of the $\mathcal{E}_{j}$'s. Under the condition (A1), the $\mathcal{E}_{j}$'s may overlap, be contained in each other, or even coincide. Generally speaking if $\max \mathcal{E}_{j} < \min\mathcal{E}_{j+1}$ for some $j=1,\ldots,\mathrm{Card}(\Theta)-1$, then we have a non-trivial spectral gap. If $\max \mathcal{E}_{j} = \min\mathcal{E}_{j+1}$, then we have a trivial spectral gap.\\
\indent Lastly, denote by $\mathcal{E}_{F}(\rho_{0})$ the Fermi energy which is defined by $\mathcal{E}_{F}(\rho_{0}) := \lim_{\beta \rightarrow \infty} \mu^{(0)}(\beta,\rho_{0},0)$. This limit always exists and actually defines an increasing function of $\rho_{0}$, see \cite[Thm. 1.1]{BCS}.\\

\indent The main results of this paper are collected in the following theorem:

\begin{theorem}
\label{thm2}
Let $\rho_{0}>0$ be fixed.\\
$\mathrm{(i)}$. For any $\beta>0$, let $\mu^{(0)}=\mu^{(0)}(\beta,\rho_{0},0) \in \mathbb{R}$ be the unique solution of the equation $\rho(\beta,\mathrm{e}^{\beta \mu},0)=\rho_{0}$. Then the bulk zero-field orbital susceptibility at fixed positive temperature and density can be written as:
\begin{equation}
\label{suscepti4}
\mathcal{X}(\beta,\rho_{0}) = \frac{1}{4} \bigg(\frac{e}{c}\bigg)^{2} \frac{1}{\mathrm{Card}(\Theta) \vert\Omega^{*}\vert} \frac{1}{\beta}  \frac{i}{2\pi} \int_{\Gamma} \mathrm{d}\xi\, \mathfrak{f}(\beta,\mu^{(0)};\xi) \int_{\Omega^{*}} \mathrm{d}\bold{k}\, \mathrm{Tr}_{l^{2}(\Theta)} \Big(\sum_{l=3}^{5} \mathcal{P}_{l}(\bold{k};\xi)\Big),
\end{equation}
where for each $\bold{k} \in \Omega^{*}$, denoting by $R(\bold{k};\xi):=(H(\bold{k}) - \xi)^{-1}$ the fiber of the resolvent:
\begin{multline}
\label{calP5}
\mathcal{P}_{5}(\bold{k};\xi) := R(\bold{k};\xi)\Big\{(\partial_{k_{1}}H)(\bold{k})R(\bold{k};\xi)(\partial_{k_{2}}H)(\bold{k}) - (\partial_{k_{2}}H)(\bold{k})R(\bold{k};\xi)(\partial_{k_{1}}H)(\bold{k})\Big\} \times \\
\times R(\bold{k};\xi)
\Big\{(\partial_{k_{2}}H)(\bold{k})R(\bold{k};\xi)(\partial_{k_{1}}H)(\bold{k}) - (\partial_{k_{1}}H)(\bold{k})R(\bold{k};\xi)(\partial_{k_{2}}H)(\bold{k})\Big\} R(\bold{k};\xi),
\end{multline}
\begin{multline}
\label{calP4}
\mathcal{P}_{4}(\bold{k};\xi) := R(\bold{k};\xi)\Big\{(\partial^{2}_{k_{1}}H)(\bold{k})R(\bold{k};\xi)(\partial_{k_{2}}H)(\bold{k}) R(\bold{k};\xi)(\partial_{k_{2}}H)(\bold{k}) + \\ - (\partial_{k_{1}}\partial_{k_{2}}H)(\bold{k})R(\bold{k};\xi)\big[(\partial_{k_{1}}H)(\bold{k}) R(\bold{k};\xi)(\partial_{k_{2}}H)(\bold{k}) + (\partial_{k_{2}}H)(\bold{k}) R(\bold{k};\xi)(\partial_{k_{1}}H)(\bold{k})\big] + \\
+ (\partial^{2}_{k_{2}}H)(\bold{k})R(\bold{k};\xi)(\partial_{k_{1}}H)(\bold{k}) R(\bold{k};\xi)(\partial_{k_{1}}H)(\bold{k})\Big\}R(\bold{k};\xi),
\end{multline}
\begin{multline}
\label{calP3}
\mathcal{P}_{3}(\bold{k};\xi) := - R(\bold{k};\xi)\Big\{ \frac{1}{2} (\partial^{2}_{k_{1}}H)(\bold{k})R(\bold{k};\xi)(\partial^{2}_{k_{2}}H)(\bold{k}) + \frac{1}{2} (\partial^{2}_{k_{2}}H)(\bold{k})R(\bold{k};\xi)(\partial^{2}_{k_{1}}H)(\bold{k}) + \\
- (\partial_{k_{1}}\partial_{k_{2}}H)(\bold{k})R(\bold{k};\xi) (\partial_{k_{1}}\partial_{k_{2}}H)(\bold{k})\Big\}R(\bold{k};\xi).
\end{multline}
$\mathrm{(ii)}$. Suppose $\mathrm{(A1)}$. Assume the semiconducting situation, i.e. the Fermi energy lies in the middle of a non-trivial gap, that is there exists $M \in \{1,\ldots, \mathrm{Card}(\Theta)-1\}$ s.t. $\max\mathcal{E}_{M}<\min\mathcal{E}_{M+1}$ and $\mathcal{E}_{F}(\rho_{0}) = (\max \mathcal{E}_{M} + \min \mathcal{E}_{M+1})/2$. Then there exist $2M$ functions $\mathfrak{d}_{j,1}(\cdot\,)$ and $\mathfrak{d}_{j,0}(\cdot\,)$, $j=1,\ldots,M$ defined on $\Omega^{*}$ outside a set of Lebesgue-measure zero s.t.
\begin{equation}
\label{xSC1}
\begin{split}
\mathcal{X}(\rho_{0}) :&= \lim_{\beta \rightarrow \infty} \mathcal{X}(\beta,\rho_{0}) \\&= \bigg(\frac{e}{c}\bigg)^{2} \frac{1}{4} \frac{1}{\mathrm{Card}(\Theta) \vert \Omega^{*}\vert} \int_{\Omega^{*}} \mathrm{d}\bold{k}\, \sum_{j=1}^{M} \Big\{ \mathfrak{d}_{j,1}(\bold{k}) + \{E_{j}(\bold{k}) - \mathcal{E}_{F}(\rho_{0})\} \mathfrak{d}_{j,0}(\bold{k})\Big\},
\end{split}
\end{equation}
and the above integrand can be extended by continuity to the whole of $\Omega^{*}$.\\
$\mathrm{(iii)}$. Suppose $\mathrm{(A2)}$ together with $\mathrm{Card}(\Theta)=2$. Assume the semiconducting situation, i.e. the Fermi energy lies in the middle of the spectral gap separating the band $\mathcal{E}_{1}$ from $\mathcal{E}_{2}$. Then in the zero-temperature limit, the bulk zero-field orbital susceptibility at fixed density is given by:
\begin{equation}
\label{xSC2}
\mathcal{X}(\rho_{0}) = \bigg(\frac{e}{c}\bigg)^{2} \frac{1}{8} \frac{1}{\vert \Omega^{*}\vert} \int_{\Omega^{*}} \mathrm{d}\bold{k}\, \sum_{l=0}^{2} \bigg\{ \frac{\mathfrak{u}_{1,l}(\bold{k})}{(E_{2}(\bold{k}) - E_{1}(\bold{k}))^{l+1}} + \frac{\{E_{1}(\bold{k}) - \mathcal{E}_{F}(\rho_{0})\} \mathfrak{v}_{1,l}(\bold{k})}{(E_{2}(\bold{k}) - E_{1}(\bold{k}))^{l+2}}\bigg\},
\end{equation}
where the functions $\mathfrak{u}_{1,l}(\cdot\,)$ and $\mathfrak{v}_{1,l}(\cdot\,)$ are respectively defined in \eqref{u10}-\eqref{u12} and \eqref{v10}-\eqref{v12}.
\end{theorem}
\vspace{0.5cm}

\noindent \textbf{Remark 1}. All functions $\mathfrak{d}_{j,1}(\cdot\,)$ and $\mathfrak{d}_{j,0}(\cdot\,)$, $j=1,\ldots,M$ appearing in \eqref{xSC1} can be explicitly written down. They only involve the eigenvalues and associated eigenfunctions of the Bloch Hamiltonian $H(\bold{k})$, together with the derivatives (up to the second order) w.r.t. the $k_{\alpha}$-variables ($\alpha=1,2$) of the $\mathrm{Card}(\Theta)\times\mathrm{Card}(\Theta)$-matrix elements of $H(\bold{k})$, see Section 3.2.1. As for the functions $\mathfrak{u}_{1,l}(\cdot\,)$ and $\mathfrak{v}_{1,l}(\cdot\,)$, $l=0,1,2$, they have the same peculiarity, see page \pageref{u10} for their explicit expressions. This special feature makes our formulas suitable for numerical computations.\\

\noindent \textbf{Remark 2}. We mention that the assumption (A1) in $\mathrm{(ii)}$ is artificial in the following sense. One actually can prove that \eqref{xSC1} still holds true when the $E_{j}$'s are degenerate on a subset of $\Omega^{*}$ with full Lebesgue-measure. However in that case the functions $\mathfrak{d}_{j,1}(\cdot\,)$ and $\mathfrak{d}_{j,0}(\cdot\,)$, $j=1,\ldots,M$ can not be expressed as it is mentioned in Remark 1: their expressions are more complicated and require the use of the orthogonal projection corresponding to each $E_{j}$. For further details, see \cite{BCS}.\\

\noindent \textbf{Remark 3}. When (A1) is replaced with (A2) in $\mathrm{(ii)}$, then \eqref{xSC1} still holds true with the same functions $\mathfrak{d}_{j,l}(\cdot\,)$, $j=1,\ldots,M$ and $l=0,1$. But in this instance these functions are smooth in the $\bold{k}$-variable by virtue of the analytic perturbation theory, see e.g. \cite[Sect. XII]{RS4}. In the particular case of $\mathrm{Card}(\Theta)=2$, one has the following identities:
\begin{equation*}
\forall\bold{k} \in \Omega^{*},\quad \mathfrak{d}_{1,1}(\bold{k}) = \sum_{l=0}^{2} \frac{\mathfrak{u}_{1,l}(\bold{k})}{(E_{2}(\bold{k}) - E_{1}(\bold{k}))^{l+1}},\quad \mathfrak{d}_{1,0}(\bold{k}) = \sum_{l=0}^{2} \frac{\mathfrak{v}_{1,l}(\bold{k})}{(E_{2}(\bold{k}) - E_{1}(\bold{k}))^{l+2}}.
\end{equation*}

\noindent \textbf{Remark 4}. As it is common practice in physics, the starting point in this kind of problem is a given matrix for the Bloch Hamiltonian $H(\bold{k})$. Since the trace in the formula \eqref{suscepti4} only involves the derivatives (up to the second order) w.r.t. the $k_{\alpha}$-variables ($\alpha=1,2$) of the $\mathrm{Card}(\Theta)\times\mathrm{Card}(\Theta)$-matrix elements of the Bloch Hamiltonian together with its 'inverse matrix', then a such formula allows 'hand'-calculations when dealing notably with $2\times2$ matrices. More precisely, the calculations from \eqref{suscepti4} are quite easy when considering dispersion laws which are polynomial in $\bold{k}$ in the gapless limit. Under these conditions, the zero-temperature limit simply follows from the formula \eqref{suscepti4} by the application of the residue theorem without resorting to the formula \eqref{xSC1} or \eqref{xSC2}, see the following paragraph for some examples. Besides we mention that under the assumption (A1), we give another formula for the zero-field orbital susceptibility at fixed positive temperature and density which have the same special feature than the one mentioned for the formula \eqref{xSC1}, see \eqref{suscepti5} in Proposition \ref{pro3.3}. In the particular case of $\mathrm{Card}(\Theta)=2$, all the terms involved in \eqref{suscepti5} are explicitly written down, see Lemma \ref{complete2}. Furthermore the assumption (A2) makes possible the rewriting of \eqref{suscepti5} as a sum of two terms: the first one is the so-called Peierls contribution, the second one stands for the interband contributions, see \eqref{exp4} in Proposition \ref{propo4}. In the zero-temperature limit and in the semiconducting situation, only a large number of the terms involved in the interband contributions will give rise to \eqref{xSC2}.\\

\noindent \textbf{Remark 5}. The regularity properties announced in $\mathrm{(ii)}$-$\mathrm{(iii)}$ of Lemma \ref{thm1} are far from being optimum. On the one hand, one can prove that $z \mapsto P(\beta,z,b)$ can be analytically extended to the complex domain $\mathbb{C}\setminus (-\infty, - \mathrm{e}^{\beta E_{b}^{(0)}}]$ where $E_{b}^{(0)} := \inf \sigma (H_{b})$, see e.g. \cite{BrCoLo1,BCS1,BS}. On the other hand, the use of the gauge invariant magnetic perturbation theory to prove $\mathrm{(iii)}$ allows us actually to get that $b \mapsto P(\beta,z,b)$ is a $\mathcal{C}^{\infty}$-function. For further details, see the proof of Proposition \ref{propo13}.

\subsection{A simple application for the two-bands gapped model - Discussions.}

The purpose of this paragraph is to present and discuss some calculations on the bulk zero-field orbital susceptibility of Bloch electrons at zero temperature and fixed density in the framework of a two-bands gapped model for which the dispersion law is linear w.r.t. the quasi-momentum in the gapless limit. For monolayer graphene, this model is a case in point as a first approximation.\\
\indent We first give a series of calculations from two distinct Bloch Hamiltonians, see \eqref{H1k} and \eqref{H2k} below, which both have the same eigenvalues. Subsequently, we discuss the results.\\

\noindent \textbf{Calculation (1).} The starting point is the following Bloch Hamiltonian:
\begin{equation}
\label{H1k}
H^{(l)}(\bold{k}) := \begin{pmatrix}
\delta & k_{1}+ik_{2} \\
k_{1} - i k_{2} & -\delta
\end{pmatrix} \quad \delta >0,\, \bold{k} \in \Omega^{*}.
\end{equation}
Let $E_{1}(\bold{k}) := - \sqrt{\delta^{2} + \vert \bold{k}\vert^{2}}$ and $E_{2}(\bold{k}):= - E_{1}(\bold{k})$ be its two eigenvalues. The matrix-resolvent is:
\begin{equation*}
R^{(l)}(\bold{k};\xi) := \frac{-1}{(E_{1}(\bold{k})-\xi)(E_{2}(\bold{k})-\xi)}
\begin{pmatrix}
\delta + \xi & k_{1}+ik_{2} \\
k_{1} - i k_{2} & \xi -\delta
\end{pmatrix} \quad \bold{k} \in \Omega^{*}.
\end{equation*}
Note that the second derivatives w.r.t. the $k_{\alpha}$-variable, $\alpha=1,2$ of the matrix elements in \eqref{H1k} are all zero. Therefore, each quantity in \eqref{calP4} and \eqref{calP3} is reduced to the $2\times 2$-zero matrix. After carrying out some calculations involving products of $2\times2$-matrices, one gets for the quantity defined in \eqref{calP5} (which we denote by $\mathcal{P}_{5}^{(l)}(\bold{k};\xi)$ to express the dependence on the model in \eqref{H1k}):
\begin{equation}
\label{calP5apl}
\mathcal{P}_{5}^{(l)}(\bold{k};\xi) := \frac{4 (\delta^{2} - \xi^{2})}{(E_{1}(\bold{k}) - \xi)^{4} (E_{2}(\bold{k}) - \xi)^{4}}\begin{pmatrix}
\delta + \xi & k_{1}+ik_{2} \\
k_{1} - i k_{2} & \xi -\delta
\end{pmatrix} \quad \bold{k} \in \Omega^{*}.
\end{equation}
Let $\beta>0$ and  $\rho_{0}>0$ be fixed.  From \eqref{calP5apl}, the formula \eqref{suscepti4} leads to:
\begin{equation*}
\mathcal{X}(\beta,\rho_{0}) = \bigg(\frac{e}{c}\bigg)^{2} \frac{1}{\vert \Omega^{*}\vert} \frac{1}{\beta} \int_{\Omega^{*}} \mathrm{d}\bold{k}\, \bigg(\frac{1}{2i\pi}\bigg) \int_{\Gamma} \mathrm{d}\xi\, \frac{\mathfrak{f}(\beta,\mu^{(0)};\xi) g(\xi)}{(E_{1}(\bold{k}) - \xi)^{4} (E_{2}(\bold{k}) -\xi)^{4}},\quad
g(\xi):= \xi(\xi^{2}-\delta^{2}).
\end{equation*}
Suppose the semiconducting situation, i.e. the Fermi energy lies in the middle of the spectral gap (the width of the gap is $2\delta$) separating the band $\mathcal{E}_{1}$ from $\mathcal{E}_{2}$. By performing first the integration w.r.t. the $\xi$-variable by the residue theorem, and subsequently the zero-temperature limit (for further details in regards to the method, see the proof of Theorem \ref{thm2} $\mathrm{(ii)}-\mathrm{(iii)}$), one has:
\begin{multline}
\label{zercont}
\mathcal{X}(\rho_{0}) := \lim_{\beta \rightarrow \infty} \mathcal{X}(\beta,\rho_{0}) = -\frac{1}{32} \bigg(\frac{e}{c}\bigg)^{2} \frac{1}{\vert \Omega^{*}\vert} \int_{\Omega^{*}} \mathrm{d}\bold{k}\, \big\{\mathfrak{d}_{1,1}(\bold{k}) + (E_{1}(\bold{k}) - \mathcal{E}_{F}(\rho_{0})) \mathfrak{d}_{1,0}(\bold{k})\big\}, \\
\textrm{where:}\quad \mathfrak{d}_{1,1}(\bold{k}) = \frac{1}{(\delta^{2} + \vert \bold{k}\vert^{2})^{\frac{3}{2}}} +  \frac{\delta^{2}}{(\delta^{2} + \vert \bold{k}\vert^{2})^{\frac{5}{2}}}\quad \textrm{and}\quad
\mathfrak{d}_{1,0}(\bold{k}) = 0.
\end{multline}
Finally, by performing first the integration w.r.t. the $\bold{k}$-variable (note that the presence of the $\delta$'s makes integrable the function $\mathfrak{d}_{1,1}(\cdot\,)$ near $\bold{k}=\bold{0}$) and only after the gapless limit, one obtains:
\begin{equation}
\label{resul1}
\lim_{\delta \rightarrow 0} \mathcal{X}(\rho_{0}) = -\infty.
\end{equation}

\noindent \textbf{Calculation (2).} The starting point is the following diagonal Bloch Hamiltonian:
\begin{equation}
\label{H2k}
H^{(d)}(\bold{k}) := \begin{pmatrix}
E_{1}(\bold{k}) & 0 \\
0 & E_{2}(\bold{k})
\end{pmatrix}, \qquad \textrm{with:}\,\,\,\left\{\begin{array}{ll} E_{1}(\bold{k}):= - \sqrt{\delta^{2} + \vert \bold{k}\vert^{2}},\\ E_{2}(\bold{k}):= - E_{1}(\bold{k})\end{array}\right. \quad \delta >0,\, \bold{k} \in \Omega^{*}.
\end{equation}
The matrix-resolvent is:
\begin{equation*}
R^{(d)}(\bold{k};\xi) := \frac{1}{(E_{1}(\bold{k})-\xi)(E_{2}(\bold{k})-\xi)} \begin{pmatrix}
E_{2}(\bold{k}) - \xi & 0 \\
0 & E_{1}(\bold{k}) - \xi
\end{pmatrix}\quad  \bold{k} \in \Omega^{*}.
\end{equation*}
Let us remark that:
\begin{equation*}
(\partial_{k_{j}}H^{(d)})(\bold{k}) := \frac{k_{j}}{\sqrt{\delta^{2} + \vert \bold{k}\vert^{2}}}
\begin{pmatrix}
- 1 & 0 \\
0 & 1
\end{pmatrix} \quad j=1,2.
\end{equation*}
Due to this feature, it is easy to see that the quantity in \eqref{calP5} is reduced to the $2\times2$-zero matrix. After working out the trace of the quantities in \eqref{calP4}-\eqref{calP3}, \eqref{suscepti4} yields:
\begin{multline*}
\mathcal{X}(\beta,\rho_{0}) = \frac{1}{8}\bigg(\frac{e}{c}\bigg)^{2} \frac{1}{\vert \Omega^{*}\vert} \frac{1}{\beta}  \sum_{j=1}^{2} \int_{\Omega^{*}} \mathrm{d}\bold{k}\, \bigg\{  \frac{(-1)^{j} \vert \bold{k}\vert^{2}}{(\delta^{2} + \vert \bold{k}\vert^{2})^{\frac{3}{2}}} \bigg(\frac{1}{2i\pi}\bigg) \int_{\Gamma} \mathrm{d}\xi\, \frac{\mathfrak{f}(\beta,\mu^{(0)};\xi)}{(E_{j}(\bold{k}) - \xi)^{4}} + \\
- \frac{\delta^{2}}{(\delta^{2} + \vert \bold{k}\vert^{2})^{2}} \bigg(\frac{1}{2i\pi}\bigg) \int_{\Gamma} \mathrm{d}\xi\, \frac{\mathfrak{f}(\beta,\mu^{(0)};\xi)}{(E_{j}(\bold{k}) - \xi)^{3}}\bigg\}.
\end{multline*}
Let us mention that the application of the residue theorem for each integral w.r.t. $\xi$ (inside the braces) provides us with a derivative of $\mathfrak{f}(\beta,\mu^{(0)},\cdot\,)$ of order three and two respectively. This feature leads, in the zero-temperature limit and in the semiconducting situation, to the following:
\begin{equation}
\label{resul2}
\mathcal{X}(\rho_{0}) := \lim_{\beta \rightarrow \infty} \mathcal{X}(\beta,\rho_{0}) = 0.
\end{equation}

Let us now discuss these results. Starting with the Bloch Hamiltonian in \eqref{H1k}, the singularity expressed as a delta function of the Fermi energy is recovered, see \eqref{resul1} and Section 1.1. Through the calculations leading to \eqref{resul2}, this singularity can be accounted for by the following.\\
\indent The Bloch Hamiltonian-matrix in \eqref{H1k} is related to the one in \eqref{H2k} by:
\begin{equation}
\label{xdcf}
H^{(l)}(\bold{k}) = P(\bold{k}) H^{(d)}(\bold{k}) P^{-1}(\bold{k}) \quad \bold{k} \in \Omega^{*},
\end{equation}
where the change-of-basis matrix together with its inverse read for instance as:
\begin{equation}
\label{changbas}
P(\bold{k}) := \begin{pmatrix}
\frac{-(k_{1} + i k_{2})}{\delta + \sqrt{\delta^{2} + \vert\bold{k}\vert^{2}}} & \frac{-(k_{1} + i k_{2})}{\delta - \sqrt{\delta^{2} + \vert\bold{k}\vert^{2}}} \\
1 & 1
\end{pmatrix},\quad P^{-1}(\bold{k}) := \begin{pmatrix}
\frac{-(k_{1} - i k_{2})}{2\sqrt{\delta^{2} + \vert \bold{k} \vert^{2}}} & \frac{\delta + \sqrt{\delta^{2} + \vert\bold{k}\vert^{2}}}{2\sqrt{\delta^{2} + \vert \bold{k} \vert^{2}}} \\
\frac{k_{1} - i k_{2}}{2\sqrt{\delta^{2} + \vert \bold{k} \vert^{2}}} & \frac{- \delta + \sqrt{\delta^{2} + \vert \bold{k} \vert^{2}}}{2\sqrt{\delta^{2} + \vert \bold{k} \vert^{2}}}
\end{pmatrix}.
\end{equation}
Denote by $\mathcal{P}_{5}^{(d)}(\bold{k};\xi)$ the quantity defined in \eqref{calP5} but with $H^{(d)}(\bold{k})$. By replacing $H^{(l)}(\bold{k})$ with the r.h.s. of \eqref{xdcf} in $\mathcal{P}_{5}^{(l)}(\bold{k};\xi)$ which only survives in the calculation (1), one gets:
\begin{equation}
\label{idfres}
\mathcal{P}_{5}^{(l)}(\bold{k};\xi) = P(\bold{k}) \mathcal{P}_{5}^{(d)}(\bold{k};\xi) P^{-1}(\bold{k}) + W_{5}^{(l,d)}(\bold{k};\xi) \quad \bold{k} \in \Omega^{*}.
\end{equation}
Here $W_{5}^{(l,d)}(\bold{k};\xi)$ consists of a sum of terms, each of them containing at least a 'derivative w.r.t. the $k_{\alpha}$-variable of the change-of-basis matrix' (or its inverse). A generical term is for example:
\begin{multline*}
P(\bold{k})R^{(d)}(\bold{k};\xi) P^{-1}(\bold{k})(\partial_{k_{j}} P)(\bold{k}) H^{(d)}(\bold{k}) R^{(d)}(\bold{k};\xi) (\partial_{k_{i}} H^{(d)})(\bold{k}) R^{(d)}(\bold{k};\xi) (\partial_{k_{i}} H^{(d)})(\bold{k})\times \\
\times R^{(d)}(\bold{k};\xi) H^{(d)}(\bold{k})(\partial_{k_{i}} P^{-1})(\bold{k})P(\bold{k})R^{(d)}(\bold{k};\xi) P^{-1}(\bold{k})\quad i,j=1,2,\, i\neq j.
\end{multline*}
As mentioned in calculation (2), $\mathcal{P}_{5}^{(d)}(\bold{k};\xi)=0$ (in the matrices-sense). In view of \eqref{idfres} and \eqref{calP5apl}, this means that only $W_{5}^{(l,d)}(\bold{k};\xi)$ (through its trace) provides a non-zero contribution to the zero-field orbital susceptibility at zero-temperature, see \eqref{zercont}. Therefore, the singular behavior appearing in the gapless limit \eqref{resul1} is closely related to the fact that the maps $\bold{k} \mapsto P(\bold{k})$ and $\bold{k} \mapsto P^{-1}(\bold{k})$ are non-differentiable in $\bold{k}=\bold{0}$ when $\delta=0$, see \eqref{changbas}.\\
\indent Finally, the identically zero result obtained in \eqref{resul2} tends to show that our 'linear model' as a first approximation for monolayer graphene is not good enough. Some numerical computations based on a more 'realistic' dispersion law derived from a 'gapped two-dimensional Wallace'-type band structure is needed.

\subsection{The content of the paper.}

Our current paper is organized as follows. Section 2 is dedicated to the proof of Lemma \ref{thm1}. The proof of $\mathrm{(i)}$ essentially is based on a simplified version of the geometric perturbation theory recently developed in \cite{CN3}. Since the method for the discrete case is fully detailed in \cite[Sect. 3.3]{BrC}, the proof of \eqref{limP} just is outlined. The crucial ingredient involved in the proof of $\mathrm{(iii)}$ is the gauge invariant magnetic perturbation theory applied to the resolvent operator $R_{b}(\xi)$. It allows us to keep a good control over the linear growth induced by the magnetic vector potential when dealing with the diagonal part of the kernel $R_{b}(\cdot\,,\cdot\,;\xi)$. Section 3 is devoted to the proof of Theorem \ref{thm2}. In Section 3.1 we focus on the bulk zero-field orbital susceptibility at fixed positive temperature and density. In particular, we prove Theorem \ref{thm2} $\mathrm{(i)}$ from the results of Lemma \ref{thm1} $\mathrm{(iii)}$ together with the Bloch-Floquet decomposition. In Sections 3.2 and 3.3 we prove Theorem \ref{thm2} $\mathrm{(ii)}$ and $\mathrm{(iii)}$ respectively. It essentially follows the outline of the proof of
\cite[Thm. 1.2 (i)]{BCS}. From the formula \eqref{suscepti4}, we separately consider both situations corresponding to each one of the assumptions (A1)-(A2) and we perform the zero-temperature limit in the semiconducting situation.

\section{Proof of Lemma \ref{thm1}.}

\subsection{Proof of $\mathrm{(i)}-\mathrm{(ii)}$.}

Under the grand-canonical conditions, let $\beta := (k_{B}T)^{-1}>0$ and $z:=\mathrm{e}^{\beta \mu} >0$. \\ For any $b \geq 0$, let $\Gamma$ be the counter-clockwise oriented simple closed contour defined by:
\begin{equation}
\label{Gamma}
\Gamma := \{ \Re \xi \in [\delta_{-},\delta_{+}],\, \Im\xi = \pm \frac{\pi}{2\beta}\} \cup \{ \Re \xi = \delta_{\pm},\,\, \Im\xi \in [-\frac{\pi}{2\beta},\frac{\pi}{2\beta}]\},
\end{equation}
where $\delta_{-}:= \inf \sigma(H_{b})- 1$, $\delta_{+}:= \sup \sigma(H_{b})+ 1$. By construction $\Gamma$ surrounds the spectrum of $H_{b}$, and the closed subset surrounding by $\Gamma$ is a strict subset of $\mathfrak{D}:=\{\zeta \in \mathbb{C}: \Im \zeta \in (-\pi/\beta, \pi/\beta)\}$ which is the holomorphic domain of the function $\xi \mapsto \mathfrak{f}(\beta,z;\xi) := \ln(1 + z\mathrm{e}^{-\beta \xi})$ for any $z>0$.\\
We point out the fact that the reals $\delta_{\pm}$ in \eqref{Gamma} can actually be chosen $b$-independent since the $l^{2}$-norm of $H_{b}$ is bounded from above by some constant uniformly in $b$. This follows from the Schur-Holmgren criterion in \cite[Eq. (1.1)]{C1} applied to the kernel \eqref{magnH_b}, combined with the estimate \eqref{esH_0}. Note also that under our conditions, there exists a constant $c=c(\beta,z)>0$ s.t.
\begin{equation}
\label{bof}
\forall\xi \in \Gamma,\quad \vert \mathfrak{f}(\beta,z;\xi)\vert \leq c \mathrm{e}^{-\beta\Re \xi}.
\end{equation}
\indent Now we use the Dunford functional calculus \cite[Sect. VI.3]{DS} to write down a more convenient formula for the grand-canonical pressure of the Bloch electrons gas in the finite-size crystal.\\
\indent First of all we need to give some properties on the resolvent operator of $H_{N,b}$ defined in \eqref{Hvfini}. Let $N \in \mathbb{N}^{*}$ and $b \geq 0$. For any $\xi \in \rho(H_{N,b})$, denote by $R_{N,b}(\xi):=(H_{N,b}-\xi)^{-1}$. Obviously $R_{N,b}(\xi)$ lives on the Hilbert space $l^{2}(\Lambda_{N})$, and moreover it owns a kernel denoted by $R_{N,b}(\cdot\,,\cdot\,;\xi):\Lambda_{N}\times\Lambda_{N} \rightarrow \mathbb{C}$. This kernel is \textit{EAD} (=Exponentially Almost Diagonal) in the following sense: for any compact subset $K$ of $\rho(H_{N,b})$, there exist two constants $c_{1},c_{2}>0$ independent of $b,N$ s.t.
\begin{equation}
\label{eadres}
\forall(\bold{x},\bold{y}) \in \Lambda_{N}^{2},\quad \sup_{\xi \in K} \vert R_{N,b}(\bold{x},\bold{y};\xi)\vert \leq c_{1} \mathrm{e}^{- c_{2} \vert \bold{x}-\bold{y}\vert}.
\end{equation}
The estimate \eqref{eadres} simply follows from the fact that the kernel of $H_{N,b}$ is \textit{EAD}, see \cite[Prop. 3.5]{BrC}.\\
An important ingredient for the following is the uniform estimate on the trace of $R_{N,b}(\xi)$:
\begin{equation}
\label{uptr}
\sup_{\xi \in K} \vert \mathrm{Tr}_{l^{2}(\Lambda_{N})} (R_{N,b}(\xi)) \vert \leq c,
\end{equation}
for some $K$-dependent constant $c>0$. To obtain \eqref{uptr}, all we have to do is write the trace as the sum over $\Lambda_{N}$ of the diagonal matrix-elements of $R_{N,b}(\xi)$, and afterwards use the estimate \eqref{eadres}.\\
\indent Let $\beta>0$, $z >0$, $b \geq 0$ and $N \in \mathbb{N}^{*}$. With $\Gamma$ the contour as in \eqref{Gamma}, introduce on $l^{2}(\Lambda_{N})$:
\begin{equation*}
\label{opcalL}
\mathcal{L}_{N}(\beta,z,b) := \frac{i}{2\pi} \int_{\Gamma} \mathrm{d}\xi\, \mathfrak{f}(\beta,z;\xi) R_{N,b}(\xi).
\end{equation*}
The Dunford functional calculus provides us with the identification $\mathcal{L}_{N}(\beta,z,b) = \ln( \mathbb{I}_{N} + z \mathrm{e}^{-\beta H_{N,b}})$ which holds on $l^{2}(\Lambda_{N})$.  In view of \eqref{pressurevf}, this allows us to define the grand-canonical pressure by:
\begin{equation}
\label{newPN}
P_{N}(\beta,z,b) = \frac{1}{\beta} \frac{1}{\mathrm{Card}(\Lambda_{N})} \mathrm{Tr} (\mathcal{L}_{N}(\beta,z,b)) = \frac{1}{\beta} \frac{1}{\mathrm{Card}(\Lambda_{N})} \frac{i}{2\pi} \int_{\Gamma} \mathrm{d}\xi\, \mathfrak{f}(\beta,z;\xi) \mathrm{Tr} (R_{N,b}(\xi)).
\end{equation}
Clearly the trace in the r.h.s. of the first equality makes sense (do not forget that we in fact deal with $N\times N$-matrices). The commutation of the complex integral w.r.t. $\xi$ with the trace in the r.h.s. of the second equality is ensured by the Tonelli's theorem, see estimates \eqref{uptr} and \eqref{bof}.\\
We draw the attention to the fact that if the following limit holds:
\begin{equation}
\label{limit}
\lim_{N \rightarrow \infty} \frac{1}{\mathrm{Card}(\Lambda_{N})} \mathrm{Tr} (R_{N,b}(\xi)) = \frac{1}{\mathrm{Card}(\Theta)} \sum_{\underline{\bold{x}} \in \Theta} R_{b}(\underline{\bold{x}},\underline{\bold{x}};\xi),
\end{equation}
then \eqref{limP} (only in the pointwise-sense) will follow from the dominated convergence theorem.\\

Let us outline the proof of \eqref{limit} which uses a simplified version of the geometric perturbation theory in  \cite[Sect. 4]{CN3}. All the details about this method in the discrete case can be found in
\cite[Sect. 3.3.1]{BrC}.
Below $\lfloor \cdot\,\rfloor$ denotes the floor function. Pick $\epsilon \in (0,1)$, and assume that $N$ is large enough. The first step consists in dividing $\Lambda_{N}$ into a 'core part' $\tilde{\Lambda}_{N}$ and an 'edge part' $\tilde{\tilde{\Lambda}}_{N}$, where:
\begin{equation*}
\tilde{\Lambda}_{N} := \{\underline{\bold{x}} + \boldsymbol{\upsilon}: \underline{\bold{x}} \in \Theta, \boldsymbol{\upsilon} = \sum_{k=1}^{2} n_{k} \bold{a}_{k}; \vert n_{k}\vert \leq N - \lfloor N^{\epsilon} \rfloor\},\quad \tilde{\tilde{\Lambda}}_{N}:= \Lambda_{N} \setminus \tilde{\Lambda}_{N}.
\end{equation*}
From a geometrical argument, one can see that $\tilde{\Lambda}_{N}$ contains $(2(N - \lfloor N^{\epsilon} \rfloor) + 1)^{2}$ W-S cells. This implies that the number of W-S cells in $\tilde{\tilde{\Lambda}}_{N}$ is of order $\mathcal{O}(N^{1+\epsilon})$ for $N$ sufficiently large.
Owing to this splitting of $\Lambda_{N}$, we expect only the contribution to $\mathrm{Tr}(R_{N,b}(\xi))$ coming from the core region $\tilde{\Lambda}_{N}$ to give rise to the limit in \eqref{limit}. With this aim in view, the geometric perturbation method consists in approximating the resolvent $R_{N,b}(\xi)$ with the operator $U_{N,b}(\xi)$ defined by:
\begin{equation}
\label{idenim}
U_{N,b}(\xi) := \chi_{\tilde{\Lambda}_{N}} R_{b}(\xi)\chi_{\tilde{\Lambda}_{N}} + \chi_{\tilde{\tilde{\Lambda}}_{N}} R_{N,b}(\xi) \chi_{\tilde{\tilde{\Lambda}}_{N}},
\end{equation}
where $\chi_{\tilde{\Lambda}_{N}}$ and $\chi_{\tilde{\tilde{\Lambda}}_{N}}$ are the characteristic functions of $\tilde{\Lambda}_{N}$ and $\tilde{\tilde{\Lambda}}_{N}$ respectively. Afterwards, it remains to control the behavior when $N \rightarrow \infty$ of the trace of the r.h.s. of \eqref{idenim}. On the one hand:
\begin{equation*}
\label{limit2}
\frac{1}{\mathrm{Card}(\Lambda_{N})} \mathrm{Tr}(\chi_{\tilde{\tilde{\Lambda}}_{N}} R_{N,b}(\xi) \chi_{\tilde{\tilde{\Lambda}}_{N}}) = \frac{1}{(2N+1)^{2}\mathrm{Card}(\Theta)} \sum_{\bold{x} \in \tilde{\tilde{\Lambda}}_{N}} R_{N,b}(\bold{x},\bold{x};\xi) = \mathcal{O}(\frac{1}{N^{1-\epsilon}}),
\end{equation*}
where we used \eqref{eadres} together with the fact that the number of W-S cells in $\tilde{\tilde{\Lambda}}_{N}$ is of order $\mathcal{O}(N^{1+\epsilon})$. On the other hand, by using that $R_{b}(\xi)$ commutes with the magnetic translations, see \eqref{transmag}:
\begin{equation*}
\label{limit3}
\frac{\mathrm{Tr}(\chi_{\tilde{\Lambda}_{N}} R_{b}(\xi) \chi_{\tilde{\Lambda}_{N}})}{\mathrm{Card}(\Lambda_{N})}  =  \frac{(2(N - \lfloor N^{\epsilon} \rfloor) + 1)^{2}}{(2 N +1)^{2} \mathrm{Card}(\Theta)} \sum_{\underline{\bold{x}} \in \Theta} R_{b}(\underline{\bold{x}},\underline{\bold{x}};\xi) = \frac{1}{\mathrm{Card}(\Theta)} \sum_{\underline{\bold{x}} \in \Theta} R_{b}(\underline{\bold{x}},\underline{\bold{x}};\xi) + \mathcal{O}(\frac{1}{N^{1-\epsilon}}).
\end{equation*}
To complete the proof of \eqref{limit}, we need to control the behavior of the trace of $\{R_{N,b}(\xi) - U_{N,b}(\xi)\}$. One can prove that $(\mathrm{Card}(\Lambda_{N}))^{-1}\mathrm{Tr}(R_{N,b}(\xi) -U_{N,b}(\xi)) = \mathcal{O}(N^{-1})$. We do not give further details.\\

In order to complete the proof of Lemma \ref{thm1} $\mathrm{(i)}$, consider for any $(\beta,z) \in [\beta_{1},\beta_{2}]\times [z_{1},z_{2}]$, with $0< \beta_{1} < \beta_{2} < \infty$ and $0<z_{1}<z_{2}<\infty$, the following quantity:
\begin{equation*}
\mathcal{Q}(\beta,z,b) :=  \frac{i}{2\pi} \int_{\Gamma} \mathrm{d}\xi\, \mathfrak{f}(\beta,z;\xi) \bigg(\frac{1}{\mathrm{Card}(\Lambda_{N})} \sum_{\bold{x} \in \Lambda_{N}} R_{N,b}(\bold{x},\bold{x};\xi) - \frac{1}{\mathrm{Card}(\Theta)} \sum_{\underline{\bold{x}} \in \Theta} R_{b}(\underline{\bold{x}},\underline{\bold{x}};\xi)\bigg),
\end{equation*}
where the $\Gamma$-contour is defined as in \eqref{Gamma} but with $\beta_{2}$ instead of $\beta$.
Then $\forall(\beta,z) \in [\beta_{1},\beta_{2}]\times [z_{1},z_{2}]$:
\begin{equation*}
\vert \mathcal{Q}(\beta,z,b) \vert \leq c \int_{\Gamma} \vert \mathrm{d}\xi\vert\, \mathrm{e}^{-\beta_{1} \Re\xi} \bigg\vert\frac{1}{\mathrm{Card}(\Lambda_{N})} \sum_{\bold{x} \in \Lambda_{N}} R_{N,b}(\bold{x},\bold{x};\xi) - \frac{1}{\mathrm{Card}(\Theta)} \sum_{\underline{\bold{x}} \in \Theta} R_{b}(\underline{\bold{x}},\underline{\bold{x}};\xi)\bigg \vert,
\end{equation*}
for some constant $c=c(\beta_{1},z_{2})>0$. By virtue of \eqref{limit}, the proof of Lemma \ref{thm1} $\mathrm{(i)}$ now is over.\\

Let us conclude this paragraph by proving $\mathrm{(ii)}$ of Lemma \ref{thm1}. Let $\beta>0$ be fixed. For any $\xi \in \mathfrak{D}$ (the holomorphic domain of $\mathfrak{f}$), $z \mapsto \mathfrak{f}(\beta,z;\xi)$ is a $\mathcal{C}^{\infty}$-function on $(0,\infty)$. In particular:
\begin{equation*}
(\partial_{z}^{m} \mathfrak{f})(\beta,z;\xi) = (m-1)! (-1)^{m+1} \frac{\mathrm{e}^{-m \beta \xi}}{(1 + z \mathrm{e}^{-\beta \xi})^{m}}\quad m \in \mathbb{N}^{*},
\end{equation*}
and $\xi \mapsto (\partial_{z}^{m} \mathfrak{f})(\beta,z;\xi)$, $m \in \mathbb{N}^{*}$ is holomorphic on $\mathfrak{D}$. Moreover it obeys the following estimate:
\begin{equation*}
\forall \xi \in \Gamma,\quad \vert (\partial_{z}^{m} \mathfrak{f})(\beta,z;\xi) \vert \leq c \mathrm{e}^{- m \beta \Re \xi}\quad m \in \mathbb{N}^{*},
\end{equation*}
for some constant $c=c(m)>0$. The proof of Lemma \ref{thm1} $\mathrm{(ii)}$ follows by standard arguments.

\subsection{Proof of $\mathrm{(iii)}$.}

The proof of $\mathrm{(iii)}$ essentially is based on the application of the so-called gauge invariant magnetic perturbation theory to the resolvent operator $R_{b}(\xi)$, see Lemma \ref{lem12} below. This method makes it possible not only to investigate the regularity of the kernel $R_{b}(\cdot\,,\cdot\,;\xi)$ w.r.t. the $b$-variable, but also to write down formulas for its partial derivatives w.r.t. $b$, see Proposition \ref{propo13}. For reader's convenience we collect in the appendix of this section all the proofs of intermediary results needed to prove $\mathrm{(iii)}$.\\
\indent For the sake of simplicity, let us see $b$ as a parameter on the whole real line. Pick a $b_{0}\in \mathbb{R}$. For any $\xi \in \rho(H_{b_{0}})$ and $b \in \mathbb{R}$, introduce on $l^{2}(\Lambda)$ the operators $\tilde{R}_{b}(\xi)$ and $\tilde{T}_{b}(\xi)$ through their kernel which respectively are defined by:
\begin{align}
\label{tildeRb}
\forall(\bold{x},\bold{y}) \in \Lambda^{2}, \quad \tilde{R}_{b}(\bold{x},\bold{y};\xi) &:= \mathrm{e}^{i \delta b \phi(\bold{x},\bold{y})} R_{b_{0}}(\bold{x},\bold{y};\xi) \quad \delta b := b-b_{0},\\
\label{tildeTb}
\tilde{T}_{b}(\bold{x},\bold{y};\xi) &:= \mathrm{e}^{i \delta b \phi(\bold{x},\bold{y})} \sum_{\bold{z} \in \Lambda} \{\mathrm{e}^{i \delta b \mathrm{fl}(\bold{x},\bold{z},\bold{y})} - 1 \} H_{b_{0}}(\bold{x},\bold{z}) R_{b_{0}}(\bold{z},\bold{y};\xi),
\end{align}
where, for any arbitrary vectors $\bold{u},\bold{v},\bold{w} \in \Lambda$, $\mathrm{fl}(\bold{u},\bold{v},\bold{w}) := \phi(\bold{u},\bold{v}) + \phi(\bold{v},\bold{w}) + \phi(\bold{w},\bold{u})$. Note that $\mathrm{fl}(\bold{u},\bold{v},\bold{w})$ stands for the flux of a unit magnetic field through the triangle generated by $\bold{u}$, $\bold{v}$, $\bold{w}$.\\
From \eqref{tildeRb}, $\tilde{R}_{b}(\xi)$ clearly is bounded on $l^{2}(\Lambda)$ since its kernel is \textit{EAD}. Ditto for $\tilde{T}_{b}(\xi)$:

\begin{lema}
\label{lem11}
Let $b_{0} \in \mathbb{R}$ be fixed. For each compact subset $K \subset \rho(H_{b_{0}})$ there exists $c_{K}>0$ s.t.:
\begin{equation}
\label{normTb}
\forall b \in \mathbb{R},\quad \sup_{\xi \in K} \Vert \tilde{T}_{b}(\xi) \Vert \leq c_{K} \vert\delta b\vert.
\end{equation}
\end{lema}

The main point of the gauge invariant magnetic perturbation theory is the below identity \eqref{opident}. It asserts that for any $b$ sufficiently close to $b_{0}$ kept fixed, the resolvent $R_{b}(\xi)$ can be approximated on $l^{2}(\Lambda)$ by $\tilde{R}_{b}(\xi)$; the $l^{2}$-norm of the 'corrective term' being of order $\mathcal{O}(\vert b-b_{0}\vert)$:

\begin{lema}
\label{lem12}
Let $b_{0} \in \mathbb{R}$ be fixed. Then for each compact subset $K \subset \rho(H_{b_{0}})$ there exists $\varsigma_{K}>0$ s.t. in the bounded operators sense on $l^{2}(\Lambda)$:
\begin{equation}
\label{opident}
\forall b \in [b_{0}-\varsigma_{K},b_{0}+\varsigma_{K}],\,\,\forall \xi \in K,\quad R_{b}(\xi) = \tilde{R}_{b}(\xi) - R_{b}(\xi) \tilde{T}_{b}(\xi).
\end{equation}
\end{lema}

The above identity, twice iterated and written in the kernels sense, allows us to establish:

\begin{proposition}
\label{propo13}
Let $b_{0} \in \mathbb{R}$ and $K\subset \rho(H_{b_{0}})$ be a compact subset. Then there exists  $\varsigma_{K}>0$ s.t. $b \mapsto R_{b}(\bold{x},\bold{y};\xi)$ is a $\mathcal{C}^{2}$-function on $(b_{0}-\varsigma_{K},b_{0}+\varsigma_{K})$ $\forall\xi \in K$ and $\forall(\bold{x},\bold{y})\in \Lambda^{2}$. In particular, its second partial derivative at $b_{0}$ reads as:
\begin{multline}
\label{derRb2}
\forall (\bold{x},\bold{y}) \in \Lambda^{2},\quad (\partial_{b}^{2} R)_{b_{0}}(\bold{x},\bold{y};\xi) = - \frac{(\phi(\bold{x},\bold{y}))^{2}}{2} R_{b_{0}}(\bold{x},\bold{y};\xi) + \\
+ \sum_{\bold{z}_{1},\bold{z}_{2} \in \Lambda} R_{b_{0}}(\bold{x},\bold{z}_{1};\xi) \{\phi(\bold{x},\bold{z}_{1}) + \phi(\bold{z}_{1},\bold{y})\} \mathrm{fl}(\bold{z}_{1},\bold{z}_{2},\bold{y}) H_{b_{0}}(\bold{z}_{1},\bold{z}_{2}) R_{b_{0}}(\bold{z}_{2},\bold{y};\xi) + \\
- \sum_{\bold{z}_{1},\ldots,\bold{z}_{4} \in \Lambda} R_{b_{0}}(\bold{x},\bold{z}_{1};\xi) \mathrm{fl}(\bold{z}_{1},\bold{z}_{2},\bold{z}_{3}) H_{b_{0}}(\bold{z}_{1},\bold{z}_{2}) R_{b_{0}}(\bold{z}_{2},\bold{z}_{3};\xi) \mathrm{fl}(\bold{z}_{3},\bold{z}_{4},\bold{y}) H_{b_{0}}(\bold{z}_{3},\bold{z}_{4}) R_{b_{0}}(\bold{z}_{4},\bold{y};\xi) + \\
+ \frac{1}{2} \sum_{\bold{z}_{1},\bold{z}_{2} \in \Lambda} R_{b_{0}}(\bold{x},\bold{z}_{1};\xi)
\{\mathrm{fl}(\bold{z}_{1},\bold{z}_{2},\bold{y})\}^{2} H_{b_{0}}(\bold{z}_{1},\bold{z}_{2}) R_{b_{0}}(\bold{z}_{2},\bold{y};\xi),
\end{multline}
and the first both terms in the r.h.s. of \eqref{derRb2} vanish when $\bold{x}=\bold{y}$ by antisymmetry of the phase.
\end{proposition}

Let us prove Lemma \ref{thm1} $\mathrm{(iii)}$. Let $\beta >0$, $z>0$ and $b_{0} \in \mathbb{R}$. With the contour $\Gamma$ defined in \eqref{Gamma} (with $b_{0}$ in place of $b$), recall that the bulk pressure of the Bloch electrons gas reads as:
\begin{equation*}
P(\beta,z,b_{0}) = \frac{1}{\beta} \frac{1}{\mathrm{Card}(\Theta)} \frac{i}{2\pi}  \int_{\Gamma} \mathrm{d}\xi\, \mathfrak{f}(\beta,z;\xi) \sum_{\underline{\bold{x}} \in \Theta} R_{b_{0}}(\underline{\bold{x}},\underline{\bold{x}};\xi).
\end{equation*}
From Proposition \ref{propo13}, we know that there exists a $\varsigma_{\Gamma}>0$ such that $\Gamma$ encloses the spectrum of $H_{b}$ for all $b$ satisfying $\vert b - b_{0}\vert < \varsigma_{\Gamma}$, and moreover $b \mapsto R_{b}(\bold{x},\bold{x};\xi)$ is a $\mathcal{C}^{2}$-function on $(b_{0}-\varsigma_{\Gamma},b_{0}+\varsigma_{\Gamma})$ $\forall \xi \in \Gamma$ and $\forall\bold{x}\in \Lambda$. On the other hand, from the expressions \eqref{derRb2} and \eqref{derRb1} (with $\bold{y}=\bold{x}$) and by using the same arguments than the ones used in the proof of Lemma \ref{lem11} below, then one has:
\begin{equation*}
\forall \bold{x}\in \Lambda,\, \forall b \in (b_{0}-\varsigma_{\Gamma},b_{0}+\varsigma_{\Gamma}),\quad  \sup_{\xi \in \Gamma} \vert (\partial_{b}^{l} R)_{b}(\bold{x},\bold{x};\xi)\vert \leq c \quad l=0,1,2,
\end{equation*}
for some $\Gamma$-dependent constant $c>0$. Lemma \ref{thm1} $\mathrm{(iii)}$ now follows by standard arguments.

\subsection{Appendix: proofs of intermediate results.}

\noindent \textbf{Proof of Lemma \ref{lem11}}. From the definition \eqref{tildeTb}, it holds on $\Lambda^{2}$:
\begin{equation*}
\vert \tilde{T}_{b}(\bold{x},\bold{y};\xi) \vert \leq \vert \delta b\vert \sum_{\bold{z}\in \Lambda} \vert \bold{x} - \bold{z}\vert \vert H_{b_{0}}(\bold{x},\bold{z})\vert \vert \bold{z} - \bold{y}\vert \vert R_{b_{0}}(\bold{z},\bold{y};\xi)\vert,
\end{equation*}
where we used that $\vert \mathrm{e}^{i x} - 1 \vert \leq \vert x \vert$ $\forall x \in \mathbb{R}$, followed by the estimate $\vert\mathrm{fl}(\bold{x},\bold{z},\bold{y})\vert \leq \vert \bold{x} - \bold{z}\vert \vert \bold{z} - \bold{y}\vert$. Now use that $H_{b_{0}}(\cdot\,,\cdot\,)$ is \textit{EAD}, and therefore $R_{b_{0}}(\cdot\,,\cdot\,;\xi)$ too by \cite[Prop. 3.5]{BrC}, to get ride of the factor $\vert \bold{x} - \bold{z}\vert$ and $\vert \bold{z} - \bold{y}\vert$ respectively. Ergo there exist two $K$-dependent constants $c_{1},c_{2} >0$ s.t.:
\begin{equation*}
\forall \xi \in K,\quad \vert \tilde{T}_{b}(\bold{x},\bold{y};\xi) \vert \leq c_{1} \vert \delta b\vert \mathrm{e}^{-c_{2} \vert \bold{x} - \bold{y}\vert}.
\end{equation*}
The lemma follows by the Shur-Holmgren criterion in \cite[Eq. (1.1)]{C1}. \qed \\

\noindent \textbf{Proof of Lemma \ref{lem12}}.
Let $K \subset \rho(H_{b_{0}})$ be a compact subset. Pick a $\xi \in K$. From \eqref{tildeRb}, one has:
\begin{equation*}
\forall (\bold{x},\bold{y}) \in \Lambda^{2},\quad (H_{b} \tilde{R}_{b}(\xi))(\bold{x},\bold{y}) = \sum_{\bold{z} \in \Lambda} \mathrm{e}^{i \delta b(\phi(\bold{x},\bold{z}) + \phi(\bold{z},\bold{y}))} \mathrm{e}^{i b_{0} \phi(\bold{x},\bold{z})} H_{0}(\bold{x},\bold{z}) R_{b_{0}}(\bold{z},\bold{y};\xi).
\end{equation*}
Note that $\phi(\bold{x},\bold{z})+\phi(\bold{z},\bold{y}) = \mathrm{fl}(\bold{x},\bold{z},\bold{y}) + \phi(\bold{x},\bold{y})$. Then from \eqref{magnH_b} together with \eqref{tildeTb}:
\begin{equation*}
(H_{b} \tilde{R}_{b}(\xi))(\bold{x},\bold{y}) = \mathrm{e}^{i \delta b \phi(\bold{x},\bold{y})} \sum_{\bold{z} \in \Lambda} H_{b_{0}}(\bold{x},\bold{z}) R_{b_{0}}(\bold{z},\bold{y};\xi) + \tilde{T}_{b}(\bold{x},\bold{y};\xi).
\end{equation*}
Since
$(H_{b_{0}} R_{b_{0}}(\xi))(\bold{x},\bold{y}) = \delta_{\bold{x},\bold{y}} + \xi R_{b_{0}}(\bold{x},\bold{y})$ with $\delta_{\bold{x},\bold{y}} := \delta_{x_{1},y_{1}} \delta_{x_{2},y_{2}}$ (the Kronecker symbol), then:
\begin{equation*}
\forall(\bold{x},\bold{y}) \in \Lambda^{2},\quad ((H_{b}- \xi) \tilde{R}_{b}(\xi))(\bold{x},\bold{y}) = \mathrm{e}^{i \delta b \phi(\bold{x},\bold{y})} \delta_{\bold{x},\bold{y}}  + \tilde{T}_{b}(\bold{x},\bold{y};\xi).
\end{equation*}
This implies in the bounded operators sense on $l^{2}(\Lambda)$:
\begin{equation}
\label{opident2}
(H_{b} - \xi) \tilde{R}_{b}(\xi) = \mathbb{I} + \tilde{T}_{b}(\xi).
\end{equation}
Next use that there exists a constant $\varsigma_{K}>0$ s.t. $K \subset \rho(H_{b})$ for all $\vert b - b_{0}\vert \leq \varsigma_{K}$, see \cite[Thm 1.1 (i)]{C1}. Thus for such $b$'s, we can invert the operator $(H_{b} - \xi)$ with $\xi \in K$ as above. The proof is over.\qed \\

\noindent \textbf{Proof of Proposition \ref{propo13}}. Let $b_{0} \in \mathbb{R}$ and $K \subset \rho(H_{b_{0}})$ be a compact subset. Pick a $\xi \in K$.\\ The starting point is the identity \eqref{opident} twice iterated. Written in the kernels sense, one has $\forall b \in \mathbb{R}$ satisfying $\vert b - b_{0}\vert \leq \varsigma_{K}$ and $\forall(\bold{x},\bold{y}) \in \Lambda^{2}$:
\begin{equation}
\label{kerident}
R_{b}(\bold{x},\bold{y};\xi) = \tilde{R}_{b}(\bold{x},\bold{y};\xi) - (\tilde{R}_{b}(\xi) \tilde{T}_{b}(\xi))(\bold{x},\bold{y}) + (\tilde{R}_{b}(\xi) \tilde{T}_{b}^{2}(\xi))(\bold{x},\bold{y}) - (R_{b}(\xi) \tilde{T}_{b}^{3}(\xi))(\bold{x},\bold{y}).
\end{equation}
For each kernel in the r.h.s. of \eqref{kerident}, we expand the exponential phase factor appearing in the kernels of $\tilde{R}_{b}(\xi)$ and $\tilde{T}_{b}(\xi)$ in Taylor series up to the second order in $\delta b$. Then we have:
\begin{equation*}
\tilde{R}_{b}(\bold{x},\bold{y};\xi) = \sum_{k=0}^{2} (\delta b)^{k} \frac{(i \phi(\bold{x},\bold{y}))^{k}}{k!} R_{b_{0}}(\bold{x},\bold{y};\xi) + \mathcal{R}_{b}^{(0)}(\bold{x},\bold{y};\xi),
\end{equation*}
\begin{multline*}
(\tilde{R}_{b}(\xi) \tilde{T}_{b}(\xi))(\bold{x},\bold{y}) = \sum_{\bold{z}_{1},\bold{z}_{2} \in \Lambda} R_{b_{0}}(\bold{x},\bold{z}_{1};\xi)\big\{i \delta b
- (\delta b)^{2}\big[(\phi(\bold{x},\bold{z}_{1})+ \phi(\bold{z}_{1},\bold{y})) + \frac{1}{2}\mathrm{fl}(\bold{z}_{1},\bold{z}_{2},\bold{y})\big]\big\} \times \\
\times \mathrm{fl}(\bold{z}_{1},\bold{z}_{2},\bold{y}) H_{b_{0}}(\bold{z}_{1},\bold{z}_{2}) R_{b_{0}}(\bold{z}_{2},\bold{y};\xi) + \mathcal{R}_{b}^{(1)}(\bold{x},\bold{y};\xi),
\end{multline*}
\begin{multline*}
(\tilde{R}_{b}(\xi) \tilde{T}_{b}^{2}(\xi))(\bold{x},\bold{y}) =  - (\delta b)^{2} \sum_{\bold{z}_{1},\ldots,\bold{z}_{4} \in \Lambda} R_{b_{0}}(\bold{x},\bold{z}_{1};\xi) \mathrm{fl}(\bold{z}_{1},\bold{z}_{2},\bold{z}_{3}) H_{b_{0}}(\bold{z}_{1},\bold{z}_{2}) R_{b_{0}}(\bold{z}_{2},\bold{z}_{3};\xi) \times \\ \times \mathrm{fl}(\bold{z}_{3},\bold{z}_{4},\bold{y}) H_{b_{0}}(\bold{z}_{3},\bold{z}_{4}) R_{b_{0}}(\bold{z}_{4},\bold{y};\xi) + \mathcal{R}_{b}^{(2)}(\bold{x},\bold{y};\xi),
\end{multline*}
where the remainder terms $\mathcal{R}_{b}^{(l)}(\bold{x},\bold{y};\xi)$, $l=0,1,2$ satisfy the property that their first two derivatives at $b_{0}$ are identically zero. Note that we can disregard the last term in the r.h.s. of \eqref{kerident} since by virtue of \eqref{normTb}, its $l^{2}$-norm behaves like $\mathcal{O}(\vert\delta b\vert^{3})$ when $\vert \delta b\vert \rightarrow 0$ uniformly in $\xi \in K$.\\
Next from the above expansions, for $b \in [b_{0}-\varsigma_{K},b_{0}+\varsigma_{K}]$ sufficiently close to $b_{0}$, it holds on $\Lambda^{2}$:
\begin{multline*}
R_{b}(\bold{x},\bold{y};\xi) - R_{b_{0}}(\bold{x},\bold{y};\xi)  =
\delta b \{i \phi(\bold{x},\bold{y}) R_{b_{0}}(\bold{x},\bold{y};\xi) + \\
 - i\sum_{\bold{z}_{1},\bold{z}_{2} \in \Lambda} R_{b_{0}}(\bold{x},\bold{z}_{1};\xi) \mathrm{fl}(\bold{z}_{1},\bold{z}_{2},\bold{y}) H_{b_{0}}(\bold{z}_{1},\bold{z}_{2}) R_{b_{0}}(\bold{z}_{2},\bold{y};\xi)\} + o(\delta b).
\end{multline*}
Perform the limit $b\rightarrow b_{0}$, and we get that the map $b\mapsto R_{b}(\bold{x},\bold{y};\xi)$ is differentiable at $b_{0}$ with:
\begin{equation}
\label{derRb1}
(\partial_{b} R)_{b_{0}}(\bold{x},\bold{y};\xi) := i \phi(\bold{x},\bold{y}) R_{b_{0}}(\bold{x},\bold{y};\xi) - i \sum_{\bold{z}_{1},\bold{z}_{2} \in \Lambda} R_{b_{0}}(\bold{x},\bold{z}_{1};\xi) \mathrm{fl}(\bold{z}_{1},\bold{z}_{2},\bold{y}) H_{b_{0}}(\bold{z}_{1},\bold{z}_{2}) R_{b_{0}}(\bold{z}_{2},\bold{y};\xi).
\end{equation}
This result can be extended on the whole of $(b_{0}-\varsigma_{K},b_{0}+\varsigma_{K})$. The proposition follows by induction.\\
Note that from \eqref{opident} iterated $n$-times and written in the kernels sense, our method allows to get the expansion of the kernel of the resolvent in power series of $\delta b$ up to the $n$-th order. By an iterating procedure, one therefore can prove that $b \mapsto R_{b}(\bold{x},\bold{y};\xi)$ is a $\mathcal{C}^{\infty}$-function near $b_{0}$, $\forall\xi \in K$ and $\forall(\bold{x},\bold{y})\in \Lambda^{2}$. Although the formula obtained for the $n$-th derivative $(\partial_{b}^{n}R)_{b_{0}}(\bold{x},\bold{y};\xi)$ is quite complicated, its diagonal part only involves the flux $\mathrm{fl}$ with the kernels $H_{0}(\cdot\,,\cdot\,)$ and $R_{0}(\cdot\,,\cdot\,;\xi)$.\qed

\section{Proof of Theorem \ref{thm2}.}

\subsection{Proof of $\mathrm{(i)}$.}

We start by writing down a general formula for the bulk orbital susceptibility under the grand-canonical conditions. From \eqref{defcalX} together with the results of Section 2.3, one directly obtains:
\begin{equation}
\label{fiform}
\mathcal{X}(\beta,z,b) = \bigg(\frac{e}{c}\bigg)^{2} \frac{1}{\beta} \frac{1}{\mathrm{Card}(\Theta)} \frac{i}{2\pi}  \int_{\Gamma} \mathrm{d}\xi\, \mathfrak{f}(\beta,z;\xi) \sum_{\underline{\bold{x}} \in \Theta} (\partial_{b}^{2}R)_{b}(\underline{\bold{x}},\underline{\bold{x}};\xi) \quad \beta>0,\, z>0,\, b \geq 0,
\end{equation}
where $(\partial_{b}^{2}R)_{b}(\cdot\,,\cdot\,;\xi)$ equals \eqref{derRb2} (only the two last terms survive for the diagonal part).\\
\indent From now on, the density of particles $\rho_{0}>0$ becomes our fixed external parameter. Under the conditions of Lemma \ref{thm1}, the relation between the fugacity and the bulk density in \eqref{rhobul} can be inverted. This is made possible since for any $\beta>0$ and $b\geq 0$, the map $z \mapsto \rho(\beta,z,b)$ is strictly increasing on $(0,\infty)$, and actually defines a $\mathcal{C}^{\infty}$-diffeomorphism of $(0,\infty)$ into itself. Then there exists a unique $z^{(0)}(\beta,\rho_{0},b) \in (0,\infty)$, and therefore a unique $\mu^{(0)}(\beta,\rho_{0},b) \in \mathbb{R}$ satisfying:
\begin{equation*}
\rho(\beta,\mathrm{e}^{\beta \mu^{(0)}(\beta,\rho_{0},b)},b) = \rho_{0} >0.
\end{equation*}
In this way, the bulk orbital susceptibility at fixed positive temperature and density is defined by:
\begin{equation}
\label{chirho}
\mathcal{X}(\beta,\rho_{0},b) := \mathcal{X}(\beta,\mu^{(0)}(\beta,\rho_{0},b),b)= \mathcal{X}(\beta,\mathrm{e}^{\beta\mu^{(0)}(\beta,\rho_{0},b)},b) \quad \beta>0,\, b\geq 0.
\end{equation}
\indent Under the zero magnetic field assumption, let us now prepare the formula \eqref{chirho} (knowing \eqref{fiform}) for the Bloch-Floquet decomposition. To do that, introduce the 'position operator' in $l^{2}(\Lambda)$ as:
\begin{equation}
\label{position}
\forall \bold{x}=(x_{1},x_{2}) \in \Lambda,\quad (\bold{X}\cdot \bold{e}_{l} \psi)(\bold{x}) = x_{l} \psi(\bold{x}) \quad \psi \in l^{2}_{c}(\Lambda),\, l =1,2,
\end{equation}
where $l_{c}^{2}(\Lambda)$ denotes the space of $l^{2}$-functions with compact support.\\
For any $\xi \in \rho(H_{0})$ and any $q,r,s,t \in \{1,2\}$, define on $l^{2}(\Lambda)$ the following families of operators:
\begin{align}
\label{calS}
\mathcal{S}_{q,r,s,t}(\xi) &:= R_{0}(\xi)[\bold{X}\cdot \bold{e}_{q},[\bold{X}\cdot \bold{e}_{r},H_{0}]][\bold{X}\cdot \bold{e}_{s},[\bold{X}\cdot \bold{e}_{t},R_{0}(\xi)]],\\
\label{calT}
\mathcal{T}_{q,r,s,t}(\xi) &:= R_{0}(\xi)[\bold{X}\cdot \bold{e}_{q},H_{0}][\bold{X}\cdot \bold{e}_{r},R_{0}(\xi)][\bold{X}\cdot \bold{e}_{s},H_{0}][\bold{X}\cdot \bold{e}_{t},R_{0}(\xi)].
\end{align}
Here $[\cdot\,,\cdot\,]$ denotes the commutator on $l^{2}(\Lambda)$. Since the kernels of $H_{0}$ and $R_{0}(\xi)$ are \textit{EAD}, then all the above commutators are well-defined on $l^{2}(\Lambda)$. As we will see hereafter, the families of operators defined in \eqref{calS} and \eqref{calT} have the feature to commute with the translations of the $\boldsymbol{\Upsilon}$-lattice.\\

\indent Involving \eqref{calS} and \eqref{calT}, the bulk zero-field orbital susceptibility at fixed positive temperature and density can be written as:

\begin{lema}
\label{pro2}
Let $\beta>0$ and $\rho_{0}>0$ be fixed. Let $\mu^{(0)}=\mu^{(0)}(\beta,\rho_{0},b=0) \in \mathbb{R}$ be the unique solution of the equation $\rho(\beta,\mathrm{e}^{\beta \mu},b=0)=\rho_{0}$. Then one has:
\begin{multline}
\label{suscepti2}
\mathcal{X}(\beta,\rho_{0},b=0)
= \frac{1}{4} \bigg(\frac{e}{c}\bigg)^{2} \frac{1}{\beta} \frac{1}{\mathrm{Card}(\Theta)} \frac{i}{2\pi} \int_{\Gamma} \mathrm{d}\xi\, \mathfrak{f}(\beta,\mu^{(0)};\xi) \sum_{\underline{\bold{x}} \in \Theta} \Big\{\frac{1}{2} \mathcal{S}_{2,2,1,1}(\xi) + \frac{1}{2} \mathcal{S}_{1,1,2,2}(\xi) + \\ - \mathcal{S}_{1,2,1,2}(\xi)
- \mathcal{T}_{2,1,2,1}(\xi) + \mathcal{T}_{2,1,1,2}(\xi) + \mathcal{T}_{1,2,2,1}(\xi) - \mathcal{T}_{1,2,1,2}(\xi)\Big\}(\underline{\bold{x}},\underline{\bold{x}}).
\end{multline}
\end{lema}

\noindent \textbf{Proof.} By inserting \eqref{derRb2} into \eqref{fiform}, then by setting $b=0$ we get:
\begin{multline}
\label{suscepti1}
\mathcal{X}(\beta,\rho_{0},b=0) =
\frac{1}{2} \bigg(\frac{e}{c}\bigg)^{2} \frac{1}{\beta} \frac{1}{\mathrm{Card}(\Theta)} \frac{i}{2\pi} \int_{\Gamma} \mathrm{d}\xi\, \mathfrak{f}(\beta,\mu^{(0)};\xi)  \times \\
\times \sum_{\underline{\bold{x}} \in \Theta} \bigg\{ \sum_{\bold{z}_{1},\bold{z}_{2} \in \Lambda} R_{0}(\underline{\bold{x}},\bold{z}_{1};\xi)\{ \mathrm{fl}(\bold{z}_{1},\bold{z}_{2},\underline{\bold{x}})\}^{2} H_{0}(\bold{z}_{1},\bold{z}_{2}) R_{0}(\bold{z}_{2},\underline{\bold{x}};\xi) + \\
- 2 \sum_{\bold{z}_{1},\ldots,\bold{z}_{4} \in \Lambda} R_{0}(\underline{\bold{x}},\bold{z}_{1};\xi) \mathrm{fl}(\bold{z}_{1},\bold{z}_{2},\bold{z}_{3})  H_{0}(\bold{z}_{1},\bold{z}_{2}) R_{0}(\bold{z}_{2},\bold{z}_{3};\xi) \mathrm{fl}(\bold{z}_{3},\bold{z}_{4},\underline{\bold{x}}) H_{0}(\bold{z}_{3},\bold{z}_{4}) R_{0}(\bold{z}_{4},\underline{\bold{x}};\xi)\bigg\}. \nonumber
\end{multline}
From the above expression, it remains to give both ingredients which lead to \eqref{suscepti2}. The first one consists in rewriting the magnetic flux $\mathrm{fl}(\bold{x},\bold{y},\bold{z})$, for arbitrary $\bold{x},\bold{y},\bold{z} \in \Lambda^{3}$, as the following:
\begin{equation*}
\mathrm{fl}(\bold{x},\bold{y},\bold{z}) := \phi(\bold{x},\bold{y}) + \phi(\bold{y},\bold{z})+\phi(\bold{z},\bold{x}) = \frac{1}{2}[(y_{1} - z_{1})(x_{2} - y_{2}) - (y_{2}-z_{2})(x_{1}-y_{1})],
\end{equation*}
where we used that $\phi(\bold{u},\bold{v}) = \frac{1}{2}(u_{2}v_{1}-u_{1}v_{2}) = \frac{1}{2}[(u_{2}-v_{2})v_{1} - (u_{1}-v_{1})v_{2}]$. The second one consists in using the following identity. Let $A(\cdot\,,\cdot\,)$ be the kernel of a bounded operator A. Then:
\begin{equation*}
\forall (\bold{x},\bold{y}) \in \Lambda^{2},\quad (x_{l} - y_{l}) A(\bold{x},\bold{y}) = [\bold{X}\cdot \bold{e}_{l},A](\bold{x},\bold{y}) \quad l =1,2,
\end{equation*}
where $\bold{X}\cdot \bold{e}_{l}$ is the position operator defined in \eqref{position}. Its kernel reads on $\Lambda^{2}$ as $(\bold{X}\cdot\bold{e}_{l})(\bold{x},\bold{y})= x_{l} \delta_{\bold{x},\bold{y}}$, $l=1,2$. The remainder of the proof only is rearranging of terms.
Note that $\mathcal{S}_{1,2,1,2}(\xi)$ in \eqref{suscepti2} can be replaced with $\mathcal{S}_{1,2,2,1}(\xi)$ or $\mathcal{S}_{2,1,1,2}(\xi)$ or $\mathcal{S}_{2,1,2,1}(\xi)$ due to the above arguments. \qed \\

It turns out that \eqref{suscepti2} actually is suitable for the Bloch-Floquet decomposition which is at the root of the formula \eqref{suscepti4}. Let us recall the framework of it, see e.g. \cite[Sect. XIII.16]{RS4}.\\
Let $\Omega^{*}$ be the Brillouin zone of the dual lattice $\boldsymbol{\Upsilon}^{*}$. Introduce the 'Bloch isometry' as:
\begin{equation*}
\begin{split}
\mathcal{U}: l_{c}^{2}(\Lambda) \mapsto \mathfrak{h} &:= \int_{\Omega^{*}}^{\oplus} \mathrm{d}\bold{k}\, l^{2}(\Theta) \\
(\mathcal{U}\psi)(\underline{\bold{x}};\bold{k}) &= \frac{1}{\vert \Omega^{*}\vert^{\frac{1}{2}}} \sum_{\boldsymbol{\upsilon} \in \boldsymbol{\Upsilon}} \mathrm{e}^{-i \bold{k}\cdot (\underline{\bold{x}} + \boldsymbol{\upsilon})} \psi(\underline{\bold{x}} + \boldsymbol{\upsilon}) \quad \bold{k} \in \Omega^{*},\,\, \underline{\bold{x}} \in \Theta,\,\, \psi \in l_{c}^{2}(\Lambda),
\end{split}
\end{equation*}
which can be extended by continuity in a unitary operator on $l^{2}(\Lambda)$. The adjoint of $\mathcal{U}$ reads as:
\begin{equation*}
(\mathcal{U}^{*}\phi)(\underline{\bold{x}}+ \boldsymbol{\upsilon}) = \frac{1}{\vert \Omega^{*}\vert^{\frac{1}{2}}} \int_{\Omega^{*}} \mathrm{d}\bold{k}\, \mathrm{e}^{i \bold{k}\cdot (\underline{\bold{x}} + \boldsymbol{\upsilon})} \phi(\underline{\bold{x}};\bold{k}) \quad \underline{\bold{x}} \in \Theta,\,\, \boldsymbol{\upsilon} \in \boldsymbol{\Upsilon},\,\, \phi(\cdot\,;\bold{k}) \in l^{2}(\Theta).
\end{equation*}
Since $H_{0}$ commutes with the translations of $\boldsymbol{\Upsilon}$, then the unitary transformation of $H_{0}$ is decomposable into a direct integral $\mathcal{U}H_{0}\mathcal{U}^{*} = \int_{\Omega^{*}}^{\oplus} \mathrm{d}\bold{k}\, H(\bold{k})$. For each $\bold{k} \in \Omega^{*}$, the fiber Hamiltonian $H(\bold{k})$ acts on $l^{2}(\Theta)$ and it is defined via its kernel which reads as, see e.g. \cite[Eq. (2.4)]{BCZ}:
\begin{equation*}
\forall(\underline{\bold{x}},\underline{\bold{y}}) \in \Theta^{2},\quad H(\underline{\bold{x}},\underline{\bold{y}};\bold{k}) := \sum_{\boldsymbol{\upsilon} \in \boldsymbol{\Upsilon}} \mathrm{e}^{-i \bold{k} \cdot (\underline{\bold{x}} + \boldsymbol{\upsilon} - \underline{\bold{y}})} H_{0}(\underline{\bold{x}} + \boldsymbol{\upsilon}, \underline{\bold{y}}) \quad \bold{k} \in \Omega^{*}.
\end{equation*}
Note that for each $\bold{k} \in \Omega^{*}$, the fiber $H(\bold{k})$ is a self-adjoint operator on $l^{2}(\Theta)$ due to the identity $H_{0}(\bold{x},\bold{y}) = \overline{H_{0}(\bold{y},\bold{x})}$ (see assumption (P1)) which implies the one $H(\underline{\bold{x}},\underline{\bold{y}};\bold{k}) = \overline{H(\underline{\bold{y}},\underline{\bold{x}};\bold{k})}$. Hereafter we denote by $R(\bold{k};\xi) := (H(\bold{k}) - \xi)^{-1}$ $\forall \xi \in \rho(H_{0})$ the fiber of the resolvent on $l^{2}(\Theta)$.\\

We now can start the actual proof of Theorem \ref{thm2} $\mathrm{(i)}$. In view of \eqref{suscepti2}, we need an expression for the fiber of the operator $\mathcal{V}_{q,r,s,t}(\xi)$, $\mathcal{V}=\mathcal{S}$ or $\mathcal{T}$. Below we use the shorthand notation $Q_{0}$ to denote $H_{0}$ or $R_{0}(\xi)$, $\xi \in \rho(H_{0})$. Due to the assumption (P3), then $[\bold{X}\cdot \bold{e}_{\alpha}, Q_{0}]$ with $\alpha=1,2$, also commute with the lattice translations. By induction, ditto for $[\bold{X}\cdot \bold{e}_{\gamma}, [\bold{X}\cdot \bold{e}_{\alpha},Q_{0}]]$ with $\alpha,\gamma =1,2$. Therefore the unitary transformation of $\mathcal{V}_{q,r,s,t}(\xi)$ is decomposable into a direct integral $\mathcal{U} \mathcal{V}_{q,r,s,t}(\xi) \mathcal{U}^{*} = \int_{\Omega^{*}}^{\oplus} \mathrm{d}\bold{k}\, \mathcal{V}_{q,r,s,t}(\bold{k};\xi)$. For each $\bold{k} \in \Omega^{*}$ the fiber $\mathcal{V}_{q,r,s,t}(\bold{k};\xi)$, $\mathcal{V}=\mathcal{S}$ or $\mathcal{T}$ lives on $l^{2}(\Theta)$ and it is defined by:
\begin{equation}
\label{fibS}
\mathcal{S}_{q,r,s,t}(\bold{k};\xi) := R(\bold{k};\xi) \big([\bold{X}\cdot \bold{e}_{q}, [\bold{X}\cdot \bold{e}_{r},H_{0}]]\big)(\bold{k}) \big([\bold{X}\cdot \bold{e}_{s}, [\bold{X}\cdot \bold{e}_{t},R_{0}(\xi)]]\big)(\bold{k}),\nonumber
\end{equation}
\begin{equation}
\label{fibT}
\mathcal{T}_{q,r,s,t}(\bold{k};\xi) := R(\bold{k};\xi) \big([\bold{X}\cdot \bold{e}_{q},H_{0}]\big)(\bold{k}) \big([\bold{X}\cdot \bold{e}_{r},R_{0}(\xi)]\big)(\bold{k})
\big([\bold{X}\cdot \bold{e}_{s},H_{0}]\big)(\bold{k}) \big([\bold{X}\cdot \bold{e}_{t},R_{0}(\xi)]\big)(\bold{k}). \nonumber
\end{equation}
Here the fibers $\big([\bold{X}\cdot \bold{e}_{\alpha}, Q_{0}]\big)(\bold{k})$ and $\big([\bold{X}\cdot \bold{e}_{\gamma}, [\bold{X}\cdot \bold{e}_{\alpha},Q_{0}]]\big)(\bold{k})$ with $Q_{0}=H_{0}$ or $R_{0}(\xi)$ and $\alpha,\gamma=1,2$ are defined through their kernel which read for each $\bold{k} \in \Omega^{*}$ and any $(\underline{\bold{x}},\underline{\bold{y}}) \in \Theta^{2}$ respectively as:
\begin{equation*}
\begin{split}
([\bold{X}\cdot \bold{e}_{\alpha}, Q_{0}])(\underline{\bold{x}},\underline{\bold{y}};\bold{k}) &:= \sum_{\boldsymbol{\upsilon} \in \boldsymbol{\Upsilon}} \mathrm{e}^{-i \bold{k} \cdot (\underline{\bold{x}} + \boldsymbol{\upsilon} - \underline{\bold{y}})} ([\bold{X}\cdot \bold{e}_{\alpha},Q_{0}])(\underline{\bold{x}} + \boldsymbol{\upsilon},\underline{\bold{y}}), \\
([\bold{X}\cdot \bold{e}_{\gamma}, [\bold{X}\cdot \bold{e}_{\alpha},Q_{0}]])(\underline{\bold{x}},\underline{\bold{y}};\bold{k}) &:= \sum_{\boldsymbol{\upsilon} \in \boldsymbol{\Upsilon}} \mathrm{e}^{-i \bold{k} \cdot (\underline{\bold{x}} + \boldsymbol{\upsilon} - \underline{\bold{y}})} ([\bold{X}\cdot \bold{e}_{\gamma}, [\bold{X}\cdot \bold{e}_{\alpha},Q_{0}]])(\underline{\bold{x}} + \boldsymbol{\upsilon},\underline{\bold{y}}).
\end{split}
\end{equation*}
Let $Q_{0}(\cdot\,,\cdot\,)$ be the kernel of $Q_{0}$. Since $([\bold{X}\cdot \bold{e}_{\alpha}, Q_{0}])(\underline{\bold{x}} + \boldsymbol{\upsilon},\underline{\bold{y}}) = (\underline{x}_{\alpha} + \upsilon_{\alpha} - \underline{y}_{\alpha}) Q_{0}(\underline{\bold{x}} + \boldsymbol{\upsilon},\underline{\bold{y}})$ and $([\bold{X}\cdot \bold{e}_{\gamma}, [\bold{X}\cdot \bold{e}_{\alpha},Q_{0}]])(\underline{\bold{x}} + \boldsymbol{\upsilon},\underline{\bold{y}}) = (\underline{x}_{\gamma} + \upsilon_{\gamma} - \underline{y}_{\gamma}) (\underline{x}_{\alpha} + \upsilon_{\alpha} - \underline{y}_{\alpha}) Q_{0}(\underline{\bold{x}} + \boldsymbol{\upsilon},\underline{\bold{y}})$, then one has on $\Theta^{2}$:
\begin{gather*}
([\bold{X}\cdot \bold{e}_{\alpha}, Q_{0}])(\underline{\bold{x}},\underline{\bold{y}};\bold{k}) = i \frac{\partial}{\partial k_{\alpha}} Q(\underline{\bold{x}},\underline{\bold{y}};\bold{k}), \quad
([\bold{X}\cdot \bold{e}_{\gamma}, [\bold{X}\cdot \bold{e}_{\alpha},Q_{0}]])(\underline{\bold{x}},\underline{\bold{y}};\bold{k}) = \big(i \frac{\partial}{\partial k_{\gamma}}\big)i \frac{\partial}{\partial k_{\alpha}} Q(\underline{\bold{x}},\underline{\bold{y}};\bold{k}), \\
\textrm{where:}\quad \forall(\underline{\bold{x}},\underline{\bold{y}}) \in \Theta^{2},\qquad Q(\underline{\bold{x}},\underline{\bold{y}};\bold{k}) := \sum_{\boldsymbol{\upsilon} \in \boldsymbol{\Upsilon}} \mathrm{e}^{-i \bold{k} \cdot (\underline{\bold{x}} + \boldsymbol{\upsilon} - \underline{\bold{y}})} Q_{0}(\underline{\bold{x}} + \boldsymbol{\upsilon},\underline{\bold{y}}) \quad \bold{k} \in \Omega^{*}.
\end{gather*}
Here these identifications are made possible since the kernels of $H_{0}$ and $R_{0}(\xi)$ are \textit{EAD}. Denoting by $(\partial_{k_{\alpha}}Q)(\bold{k})$ and $(\partial_{k_{\gamma}} \partial_{k_{\alpha}} Q)(\bold{k})$ the operators on $l^{2}(\Theta)$ generated respectively by the kernels:
\begin{equation*}
(\partial_{k_{\alpha}}Q)(\underline{\bold{x}},\underline{\bold{y}};\bold{k}) := \frac{\partial}{\partial k_{\alpha}} Q(\underline{\bold{x}},\underline{\bold{y}};\bold{k}), \quad
(\partial_{k_{\gamma}} \partial_{k_{\alpha}}Q)(\underline{\bold{x}},\underline{\bold{y}};\bold{k}) := \frac{\partial^{2}}{\partial k_{\gamma} \partial_{k_{\alpha}}} Q(\underline{\bold{x}},\underline{\bold{y}};\bold{k})= \frac{\partial^{2}}{\partial k_{\alpha} \partial_{k_{\gamma}}} Q(\underline{\bold{x}},\underline{\bold{y}};\bold{k}),
\end{equation*}
then for each $\bold{k} \in \Omega^{*}$ the fiber $\mathcal{V}_{q,r,s,t}(\bold{k};\xi)$, $\mathcal{V}=\mathcal{S}$ or $\mathcal{T}$ can be rewritten as:
\begin{gather}
\label{Scalk}
\mathcal{S}_{q,r,s,t}(\bold{k};\xi) = R(\bold{k};\xi)(\partial_{k_{q}} \partial_{k_{r}} H)(\bold{k})(\partial_{k_{s}} \partial_{k_{t}} R)(\bold{k};\xi),\\
\label{Tcalk}
\mathcal{T}_{q,r,s,t}(\bold{k};\xi) = R(\bold{k};\xi)(\partial_{k_{q}}H)(\bold{k})(\partial_{k_{r}} R)(\bold{k};\xi)(\partial_{k_{s}}H)(\bold{k})(\partial_{k_{t}} R)(\bold{k};\xi).
\end{gather}
Since the kernel of $\mathcal{V}_{q,r,s,t}(\xi)$ is related to the one of the fiber $\mathcal{V}_{q,r,s,t}(\bold{k};\xi)$ by:
\begin{equation*}
\forall (\bold{x},\bold{y}) \in \Lambda^{2},\quad (\mathcal{V}_{q,r,s,t}(\xi))(\bold{x},\bold{y}) = \frac{1}{\vert \Omega^{*}\vert} \int_{\Omega^{*}} \mathrm{d}\bold{k}\, \mathrm{e}^{i \bold{k} \cdot (\bold{x} - \bold{y})} (\mathcal{V}_{q,r,s,t}(\bold{k};\xi))(\bold{x},\bold{y})\quad \textrm{$\mathcal{V}=\mathcal{S}$ or $\mathcal{T}$},
\end{equation*}
where we have used \cite[Eq. (4.1)]{CN}, then the expression in \eqref{suscepti2} can be rewritten as:
\begin{multline*}
\mathcal{X}(\beta,\rho_{0},0) =  \frac{1}{4} \bigg(\frac{e}{c}\bigg)^{2}\frac{1}{\beta} \frac{1}{\mathrm{Card}(\Theta) \vert \Omega^{*}\vert} \frac{i}{2\pi} \int_{\Gamma} \mathrm{d}\xi\, \mathfrak{f}(\beta,\mu^{(0)};\xi) \int_{\Omega^{*}} \mathrm{d}\bold{k}\, \sum_{\underline{\bold{x}} \in \Theta} \Big\{\frac{1}{2} \mathcal{S}_{2,2,1,1}(\bold{k};\xi) + \\
+ \frac{1}{2} \mathcal{S}_{1,1,2,2}(\bold{k};\xi)
- \mathcal{S}_{1,2,1,2}(\bold{k};\xi)
- \mathcal{T}_{2,1,2,1}(\bold{k};\xi) + \mathcal{T}_{2,1,1,2}(\bold{k};\xi) + \mathcal{T}_{1,2,2,1}(\bold{k};\xi) - \mathcal{T}_{1,2,1,2}(\bold{k};\xi)\Big\}(\underline{\bold{x}},\underline{\bold{x}}).
\end{multline*}
Next it remains to insert in \eqref{Scalk} and \eqref{Tcalk} these identities:
\begin{gather*}
(\partial_{k_{\alpha}} R)(\bold{k};\xi) = - R(\bold{k};\xi) (\partial_{k_{\alpha}} H)(\bold{k}) R(\bold{k};\xi) \quad \bold{k} \in \Omega^{*},\, \alpha =1,2,\\
\begin{split}
(\partial_{k_{\gamma}} \partial_{k_{\alpha}}R)(\bold{k};\xi) &= R(\bold{k};\xi) (\partial_{k_{\gamma}} H)(\bold{k}) R(\bold{k};\xi) (\partial_{k_{\alpha}} H)(\bold{k}) R(\bold{k};\xi) - R(\bold{k};\xi) (\partial_{k_{\gamma}} \partial_{k_{\alpha}} H)(\bold{k}) R(\bold{k};\xi) + \\
& + R(\bold{k};\xi) (\partial_{k_{\alpha}} H)(\bold{k}) R(\bold{k};\xi) (\partial_{k_{\gamma}} H)(\bold{k}) R(\bold{k};\xi)  \quad \bold{k} \in \Omega^{*},\, \alpha,\gamma = 1,2.
\end{split}
\end{gather*}
Thus \eqref{suscepti4} follows by gathering together the terms by decreasing number of fiber of resolvent.

\subsection{Proof of $\mathrm{(ii)}-\mathrm{(iii)}$.}

The starting point is the expression of the bulk zero-field orbital susceptibility at fixed positive temperature and density obtained in \eqref{suscepti4}. In view of \eqref{xSC1} we need to subject this formula to some transformations again so as to perform the integration w.r.t. the $\xi$-variable. Only after we will perform the zero-temperature limit. Let us remark that the presence of fiber resolvents in \eqref{suscepti4} indicates the possibility to bring into play, through their kernel, the eigenvalues and associated eigenfunctions of the fiber Hamiltonian. As we will see below, this will turn out to be essential to perform the integral w.r.t. the $\xi$-variable by the residue theorem.\\
\indent Before beginning, let us fix notation. For each $\bold{k} \in \Omega^{*}$, denote by $\{E_{j}(\bold{k})\}_{j=1}^{\mathrm{Card}(\Theta)}$ the set of eigenvalues of the fiber Hamiltonian $H(\bold{k})$ counting multiplicities and in increasing order. Due to this choice of indexation, the $E_{j}$'s are $\boldsymbol{\Upsilon}^{*}$-periodic and continuous functions, but they are not differentiable at (possible) crossing-points. Denote by $\{u_{j}(\cdot\,;\bold{k})\}_{j=1}^{\mathrm{Card}(\Theta)}$ the set of associated eigenfunctions. They form a complete orthonormal system in $l^{2}(\Theta)$. Unlike the $E_{j}$'s, the eigenfunctions $\bold{k} \mapsto u_{j}(\cdot\,;\bold{k})$ may be not continuous at (possible) crossing-points since they are defined up to a $\bold{k}$-dependent phase factor which cannot be always chosen to be continuous at crossing points.\\
\indent For any integer $l,m =1,\ldots,\mathrm{Card}(\Theta)$ and $\alpha,\gamma =1,2$, let us define the quantities:
\begin{gather}
\label{defpi}
\hat{\pi}_{l,m}(\alpha;\bold{k}) := \sum_{\underline{\bold{x}} \in \Theta} \overline{u_{l}(\underline{\bold{x}};\bold{k})}[(\partial_{k_{\alpha}} H)(\bold{k}) u_{m}(\underline{\bold{x}};\bold{k})]\quad \bold{k} \in \Omega^{*},\\
\label{defsigma}
\hat{\sigma}_{l,m}(\alpha,\gamma;\bold{k}) := \sum_{\underline{\bold{x}} \in \Theta} \overline{u_{l}(\underline{\bold{x}};\bold{k})}[(\partial_{k_{\alpha}} \partial_{k_{\gamma}}H)(\bold{k}) u_{m}(\underline{\bold{x}};\bold{k})]\quad \bold{k} \in \Omega^{*}.
\end{gather}
Due to the phase factors presence in the $u_{j}$'s, both above quantities may be not continuous at (possible) crossing points. Besides from \eqref{defpi}, one has $\hat{\pi}_{l,m}(\alpha;\bold{k}) = \overline{\hat{\pi}_{m,l}(\alpha;\bold{k})}$ since:
\begin{equation*}
\sum_{\underline{\bold{x}} \in \Theta} \overline{u_{l}(\underline{\bold{x}};\bold{k})} \sum_{\underline{\bold{y}} \in \Theta} \frac{\partial}{\partial k_{\alpha}} H(\underline{\bold{x}},\underline{\bold{y}};\bold{k}) u_{m}(\underline{\bold{y}};\bold{k})  = \sum_{\underline{\bold{y}} \in \Theta} u_{m}(\underline{\bold{y}};\bold{k}) \overline{\sum_{\underline{\bold{x}} \in \Theta} \frac{\partial}{\partial k_{\alpha}} H(\underline{\bold{y}},\underline{\bold{x}};\bold{k}) u_{l}(\underline{\bold{x}};\bold{k})},
\end{equation*}
where we used that $H(\underline{\bold{x}},\underline{\bold{y}};\bold{k}) = \overline{H(\underline{\bold{y}},\underline{\bold{x}};\bold{k})}$ $\forall(\underline{\bold{x}},\underline{\bold{y}}) \in \Theta^{2}$. The same property holds for $\hat{\sigma}_{l,m}(\alpha,\gamma;\bold{k})$.\\

\indent The sequel of this paragraph is divided into two parts. In the first one, we treat the generical situation in which the $E_{j}(\cdot\,)$'s are non-degenerate outside a subset of $\Omega^{*}$ with Lebesgue-measure zero. In the second one, we restrict the number of sites in $\Theta$ to two sites and we consider the situation in which both $E_{1}(\cdot\,)$ and $E_{2}(\cdot\,)$ are non-degenerate everywhere on $\Omega^{*}$.

\subsubsection{Proof of $\mathrm{(ii)}$.}

In what follows we tacitly suppose the assumption (A1), see page \pageref{x}.\\
Here is the last rewriting of the zero-field orbital susceptibility before the zero-temperature limit:

\begin{proposition}
\label{pro3.3}
Let $\beta>0$ and $\rho_{0}>0$ be fixed. Let $\mu^{(0)}=\mu^{(0)}(\beta,\rho_{0},0) \in \mathbb{R}$ be the unique solution of the equation $\rho(\beta,\mathrm{e}^{\beta \mu},0)=\rho_{0}$. Then for each integer $j= 1,\ldots,\mathrm{Card}(\Theta)$ there exist four families of functions $\mathfrak{d}_{j,l}(\cdot\,)$ with $l=0,1,2,3$ defined on $\Omega^{*}$ outside a set of Lebesgue-measure zero, s.t. the integrand below can be extended by continuity to the whole of $\Omega^{*}$, and:
\begin{equation}
\label{suscepti5}
\mathcal{X}(\beta,\rho_{0},0) = -\frac{1}{4} \bigg(\frac{e}{c}\bigg)^{2} \frac{1}{\mathrm{Card}(\Theta) \vert\Omega^{*}\vert} \frac{1}{\beta} \sum_{j=1}^{\mathrm{Card}(\Theta)} \int_{\Omega^{*}} \mathrm{d}\bold{k}\,  \sum_{l=0}^{3} \frac{\partial^{l} \mathfrak{f}}{\partial \xi^{l}}(\beta,\mu^{(0)};E_{j}(\bold{k})) \mathfrak{d}_{j,l}(\bold{k}),
\end{equation}
with the convention $(\partial_{\xi}^{0} \mathfrak{f})(\beta,\mu^{(0)};E_{j}(\bold{k})):= \mathfrak{f}(\beta,\mu^{(0)};E_{j}(\bold{k}))= \ln(1 + \mathrm{e}^{\beta(\mu^{(0)} - E_{j}(\bold{k}))})$.
\end{proposition}

The special feature of this formula lies in the fact that each function $\mathfrak{d}_{j,l}(\cdot\,)$ can be only expressed in terms of eigenvalues and associated eigenfunctions of the Bloch Hamiltonian $H(\bold{k})$, together with the derivatives (up to the second order) w.r.t. the $k_{\alpha}$-variables ($\alpha=1,2$) of the $\mathrm{Card}(\Theta)\times\mathrm{Card}(\Theta)$-matrix elements of $H(\bold{k})$. By way of example, the functions $\mathfrak{d}_{j,3}(\cdot\,)$, $j=1,\ldots,\mathrm{Card}(\Theta)$ are identified in \eqref{dtroi}. Note that all functions $\mathfrak{d}_{j,l}(\cdot\,)$, $l=0,1,2$ can also be written down, but their explicit expression is not necessary for the proof of Theorem \ref{thm2} $\mathrm{(ii)}$. Further, the expansion obtained in \eqref{suscepti5} makes the choice of phase for the $u_{j}(\cdot\,;\bold{k})$'s irrelevant.\\

Now we give the proof of Proposition \ref{pro3.3}; it follows the outline of the proof of \cite[Thm 3.1]{BCS}.
For reader's convenience we collect in the appendix of this section the proofs of intermediary results.\\
For the sake of simplicity, let us rewrite the zero-field orbital susceptibility formula in \eqref{suscepti4} as:
\begin{gather}
\label{suscepti4'}
\mathcal{X}(\beta,\rho_{0},0) = \frac{1}{4} \bigg(\frac{e}{c}\bigg)^{2} \frac{1}{\mathrm{Card}(\Theta) \vert \Omega^{*}\vert} \frac{1}{\beta} \sum_{l=3}^{5} \mathcal{W}_{l}(\beta,\mu^{(0)},0),\\
\textrm{where:}\quad \mathcal{W}_{l}(\beta,\mu^{(0)},0) := \frac{i}{2\pi} \int_{\Gamma} \mathrm{d}\xi\, \mathfrak{f}(\beta,\mu^{(0)};\xi) \int_{\Omega^{*}} \mathrm{d}\bold{k}\, \mathrm{Tr}_{l^{2}(\Theta)} (\mathcal{P}_{l}(\bold{k};\xi)) \quad l=3,4,5. \nonumber
\end{gather}

The first step consists in expressing the kernel of each fiber resolvent appearing in $\mathcal{P}_{l}(\bold{k};\xi)$, $l=3,4,5$ (see formulas \eqref{calP3}-\eqref{calP5}) in terms of eigenvalues and eigenfunctions of the fiber Hamiltonian (via the spectral theorem). Thus the quantities $\mathcal{W}_{l}(\beta,\mu^{(0)},0)$ can be rewritten as:

\begin{lema}
\label{lemm1}
Let $\beta>0$ and $\rho_{0}>0$ be fixed. Then one has:
\begin{multline}
\label{trP5}
\mathcal{W}_{5}(\beta,\mu^{(0)},0) =
-  \sum_{j_{1},\ldots,j_{4}=1}^{\mathrm{Card}(\Theta)} \int_{\Omega^{*}} \mathrm{d}\bold{k}\,  \mathcal{C}_{j_{1},j_{2},j_{3},j_{4}}(\bold{k}) \times \\
\times \frac{1}{2i \pi} \int_{\Gamma} \mathrm{d}\xi\, \frac{\mathfrak{f}(\beta,\mu^{(0)};\xi)}{(E_{j_{1}}(\bold{k}) - \xi)^{2} (E_{j_{2}}(\bold{k}) - \xi) (E_{j_{3}}(\bold{k}) - \xi) (E_{j_{4}}(\bold{k}) - \xi)},
\end{multline}
\begin{multline}
\label{trP4}
\mathcal{W}_{4}(\beta,\mu^{(0)},0) =
-  \sum_{j_{1},j_{2},j_{3}=1}^{\mathrm{Card}(\Theta)} \int_{\Omega^{*}} \mathrm{d}\bold{k}\,  \mathcal{C}_{j_{1},j_{2},j_{3}}(\bold{k}) \times \\
\times \frac{1}{2i \pi} \int_{\Gamma} \mathrm{d}\xi\, \frac{\mathfrak{f}(\beta,\mu^{(0)};\xi)}{(E_{j_{1}}(\bold{k}) - \xi)^{2} (E_{j_{2}}(\bold{k}) - \xi) (E_{j_{3}}(\bold{k}) - \xi)},
\end{multline}
\begin{equation}
\label{trP3}
\mathcal{W}_{3}(\beta,\mu^{(0)},0) =
-  \sum_{j_{1},j_{2}=1}^{\mathrm{Card}(\Theta)} \int_{\Omega^{*}} \mathrm{d}\bold{k}\,  \mathcal{C}_{j_{1},j_{2}}(\bold{k}) \frac{1}{2i \pi} \int_{\Gamma} \mathrm{d}\xi\, \frac{\mathfrak{f}(\beta,\mu^{(0)};\xi)}{(E_{j_{1}}(\bold{k}) - \xi)^{2} (E_{j_{2}}(\bold{k}) - \xi)},
\end{equation}
where $\forall\bold{k} \in \Omega^{*}$, the functions $\mathcal{C}_{j_{1},j_{2},j_{3},j_{4}}(\cdot\,)$, $\mathcal{C}_{j_{1},j_{2},j_{3}}(\cdot\,)$ and $\mathcal{C}_{j_{1},j_{2}}(\cdot\,)$ are respectively defined by:
\begin{multline}
\label{C4}
\mathcal{C}_{j_{1},j_{2},j_{3},j_{4}}(\bold{k}) := \{ \hat{\pi}_{j_{1},j_{2}}(1;\bold{k}) \hat{\pi}_{j_{2},j_{3}}(2;\bold{k}) - \hat{\pi}_{j_{1},j_{2}}(2;\bold{k}) \hat{\pi}_{j_{2},j_{3}}(1;\bold{k})\} \times \\
 \times \{ \hat{\pi}_{j_{3},j_{4}}(2;\bold{k}) \hat{\pi}_{j_{4},j_{1}}(1;\bold{k}) - \hat{\pi}_{j_{3},j_{4}}(1;\bold{k}) \hat{\pi}_{j_{4},j_{1}}(2;\bold{k})\},
\end{multline}
\begin{multline}
\label{C3}
\mathcal{C}_{j_{1},j_{2},j_{3}}(\bold{k}) := \hat{\sigma}_{j_{1},j_{2}}(1,1;\bold{k}) \hat{\pi}_{j_{2},j_{3}}(2;\bold{k})\hat{\pi}_{j_{3},j_{1}}(2;\bold{k}) + \hat{\sigma}_{j_{1},j_{2}}(2,2;\bold{k}) \hat{\pi}_{j_{2},j_{3}}(1;\bold{k})\hat{\pi}_{j_{3},j_{1}}(1;\bold{k}) + \\
- \hat{\sigma}_{j_{1},j_{2}}(1,2;\bold{k}) \hat{\pi}_{j_{2},j_{3}}(1;\bold{k})\hat{\pi}_{j_{3},j_{1}}(2;\bold{k}) - \hat{\sigma}_{j_{1},j_{2}}(1,2;\bold{k}) \hat{\pi}_{j_{2},j_{3}}(2;\bold{k})\hat{\pi}_{j_{3},j_{1}}(1;\bold{k}),
\end{multline}
\begin{equation}
\label{C2}
\mathcal{C}_{j_{1},j_{2}}(\bold{k}) := - \Re \{\hat{\sigma}_{j_{1},j_{2}}(1,1;\bold{k}) \hat{\sigma}_{j_{2},j_{1}}(2,2;\bold{k})\}
+ \hat{\sigma}_{j_{1},j_{2}}(1,2;\bold{k}) \hat{\sigma}_{j_{2},j_{1}}(1,2;\bold{k}).
\end{equation}
The quantities $\hat{\pi}_{l,m}(\alpha;\bold{k})$ and $\hat{\sigma}_{l,m}(\alpha,\gamma;\bold{k})$, $\alpha,\gamma=1,2$ are respectively defined in \eqref{defpi} and \eqref{defsigma}.
\end{lema}

The second step consists in applying the residue theorem in \eqref{trP5}-\eqref{trP3}. It straightforwardly provides us with:

\begin{lema}
\label{lemm2}
For each integer $j_{1}=1,\ldots,\mathrm{Card}(\Theta)$, there exist four families of functions $\mathfrak{a}_{j_{1},l}(\cdot\,)$ with $l=0,1,2,3$ defined on $\Omega^{*}$ outside a set of Lebesgue-measure zero s.t.:
\begin{equation}
\label{exp5}
\mathcal{W}_{5}(\beta,\mu^{(0)},0) = - \sum_{j_{1}=1}^{\mathrm{Card}(\Theta)} \int_{\Omega^{*}} \mathrm{d}\bold{k}\, \sum_{l=0}^{3} \frac{\partial^{l} \mathfrak{f}}{\partial \xi^{l}}(\beta,\mu^{(0)};E_{j_{1}}(\bold{k})) \mathfrak{a}_{j_{1},l}(\bold{k}).
\end{equation}
In particular, for Lebesgue-almost all $\bold{k} \in \Omega^{*}$ the function $\mathfrak{a}_{j_{1},3}(\cdot\,)$ is defined by:
\begin{equation*}
\mathfrak{a}_{j_{1},3}(\bold{k}) := \frac{1}{3!} \sum_{\substack{j_{2}=1 \\ j_{2} \neq j_{1}}}^{\mathrm{Card}(\Theta)} \frac{\mathcal{C}_{j_{1},j_{1},j_{2},j_{1}}(\bold{k})}{E_{j_{2}}(\bold{k}) - E_{j_{1}}(\bold{k})}.
\end{equation*}
\end{lema}

\begin{lema}
\label{lemm3}
For each integer $j_{1}=1,\ldots,\mathrm{Card}(\Theta)$, there exist four families of functions $\mathfrak{b}_{j_{1},l}(\cdot\,)$ with $l=0,1,2,3$ defined on $\Omega^{*}$ outside a set of Lebesgue-measure zero s.t.:
\begin{equation}
\label{exp4}
\mathcal{W}_{4}(\beta,\mu^{(0)},0) = - \sum_{j_{1}=1}^{\mathrm{Card}(\Theta)} \int_{\Omega^{*}} \mathrm{d}\bold{k}\, \sum_{l=0}^{3} \frac{\partial^{l} \mathfrak{f}}{\partial \xi^{l}}(\beta,\mu^{(0)};E_{j_{1}}(\bold{k})) \mathfrak{b}_{j_{1},l}(\bold{k}).
\end{equation}
In particular, for (Lebesgue-almost) all $\bold{k} \in \Omega^{*}$ the function $\mathfrak{b}_{j_{1},3}(\cdot\,)$ is defined by:
\begin{equation*}
\mathfrak{b}_{j_{1},3}(\bold{k}) := \frac{1}{3!} \mathcal{C}_{j_{1},j_{1},j_{1}}(\bold{k}).
\end{equation*}
\end{lema}

\begin{lema}
\label{lemm4}
For each integer $j_{1}=1,\ldots,\mathrm{Card}(\Theta)$, there exist three families of functions $\mathfrak{c}_{j_{1},l}(\cdot\,)$ with $l=0,1,2$ defined on $\Omega^{*}$ outside a set of Lebesgue-measure zero s.t.:
\begin{equation}
\label{exp3}
\mathcal{W}_{3}(\beta,\mu^{(0)},0) = - \sum_{j_{1}=1}^{\mathrm{Card}(\Theta)} \int_{\Omega^{*}} \mathrm{d}\bold{k}\, \sum_{l=0}^{2} \frac{\partial^{l} \mathfrak{f}}{\partial \xi^{l}}(\beta,\mu^{(0)};E_{j_{1}}(\bold{k})) \mathfrak{c}_{j_{1},l}(\bold{k}).
\end{equation}
\end{lema}

We stress the point that all functions appearing in the expansions \eqref{exp5}-\eqref{exp3} can be written down since a such result only is based on identities provided by the residue theorem. This will be done in the second part of this paragraph when considering the case of $\mathrm{Card}(\Theta)=2$.\\
\indent Thus Lemmas \ref{lemm2}, \ref{lemm3} and \ref{lemm4} provide us with an expansion of the type announced
in \eqref{suscepti5}, where the coefficients read for $j_{1}=1,\ldots,\mathrm{Card}(\Theta)$ and almost everywhere on $\Omega^{*}$ as:
\begin{gather}
\label{dtroi}
\mathfrak{d}_{j_{1},3}(\bold{k}) :=  \frac{1}{3!} \bigg\{ \sum_{\substack{j_{2} =1 \\ j_{2} \neq j_{1}}}^{\mathrm{Card}(\Theta)} \frac{\mathcal{C}_{j_{1},j_{1},j_{2},j_{1}}(\bold{k})}{E_{j_{2}}(\bold{k}) - E_{j_{1}}(\bold{k})} + \mathcal{C}_{j_{1},j_{1},j_{1}}(\bold{k})\bigg\},\\
\label{dtroi2}
\mathfrak{d}_{j_{1},l}(\bold{k}) := \mathfrak{a}_{j_{1},l}(\bold{k}) + \mathfrak{b}_{j_{1},l}(\bold{k}) + \mathfrak{c}_{j_{1},l}(\bold{k}) \quad l=0,1,2.
\end{gather}

To conclude the proof of Proposition \ref{pro3.3}, we need the following result:
\begin{lema}
\label{lemm5}
The functions $\mathfrak{a}_{j_{1},l}(\cdot\,)$, $\mathfrak{b}_{j_{1},l}(\cdot\,)$ with $l=0,1,2,3$ and $\mathfrak{c}_{j_{1},l}(\cdot\,)$, $l=0,1,2$ are bounded and continuous on any compact subset of $\Omega^{*}$ where $E_{j_{1}}$ is isolated from the rest of the spectrum.
\end{lema}

We do not give a proof of Lemma \ref{lemm5} since all details can be found in the proof of \cite[Lem. 3.7]{BCS}. Thereby the functions $\mathfrak{d}_{j_{1},l}(\cdot\,)$ in \eqref{dtroi}-\eqref{dtroi2} might be singular on a set of Lebesgue-measure zero where $E_{j_{1}}$ can touch the neighboring bands. However the whole integrand in \eqref{suscepti5} comes from the complex integrals \eqref{trP5}-\eqref{trP3} which have no local singularities in $\bold{k}$ (they actually are bounded uniformly in $\bold{k}$, see the proof of Lemma \ref{lemm1}). The proof of Proposition \ref{pro3.3} is now over.\\

We now can start the actual proof of Theorem \ref{thm2} $\mathrm{(ii)}$; it follows the outline of
\cite[Sect. 4.1]{BCS}.\\
Let $\rho_{0}>0$ be fixed. Let $\mu^{(0)} = \mu^{(0)}(\beta,\rho_{0},0) \in \mathbb{R}$ be the unique solution of $\rho(\beta,\mathrm{e}^{\beta \mu},0)=\rho_{0}$. Let $\mathfrak{f}_{FD}(\beta,\mu^{(0)};\xi) := (\mathrm{e}^{\beta (\xi - \mu^{(0)})} + 1)^{-1} = - \beta^{-1} (\partial_{\xi} \mathfrak{f})(\beta,\mu^{(0)};\xi)$ be the Fermi-Dirac (F-D) distribution.\\
By involving the F-D distribution and its derivatives, the expansion \eqref{suscepti5} can be rewritten as:
\begin{multline}
\label{dernt}
\mathcal{X}(\beta,\rho_{0},0) = \frac{1}{4} \bigg(\frac{e}{c}\bigg)^{2} \frac{1}{\mathrm{Card(\Theta)} \vert \Omega^{*}\vert} \sum_{j=1}^{\mathrm{Card}(\Theta)} \int_{\Omega^{*}} \mathrm{d}\bold{k}\, \bigg\{ \sum_{l=0}^{2} \frac{\partial^{l} \mathfrak{f}_{FD}}{\partial \xi^{l}}(\beta,\mu^{(0)};E_{j}(\bold{k})) \mathfrak{d}_{j,l+1}(\bold{k}) + \\
- \frac{1}{\beta} \mathfrak{f}(\beta,\mu^{(0)};E_{j}(\bold{k})) \mathfrak{d}_{j,0}(\bold{k})\bigg\}.
\end{multline}
Consider the semiconducting situation, i.e. assume that the Fermi energy lies in the middle of a non-trivial gap. Then there exists $M \in \{1,\ldots,\mathrm{Card}(\Theta)-1\}$ such that $\max \mathcal{E}_{M} < \min \mathcal{E}_{M+1}$, and moreover, $\lim_{\beta \rightarrow \infty} \mu^{(0)}(\beta,\rho_{0}) = (\max \mathcal{E}_{M} + \min \mathcal{E}_{M+1})/2$. From the pointwise convergences \cite[Eqs. (4.2)-(4.3)]{BCS} together with the estimates \cite[Eq. (4.4)]{BCS}, then all terms in \eqref{dernt} containing derivatives of the Fermi-Dirac distribution will converge to zero in the limit $\beta \rightarrow \infty$. This leads to \eqref{xSC1}.

\subsubsection{Proof of $\mathrm{(iii)}$.}

In this paragraph, we implicitly suppose the assumption (A2), see page \pageref{y}.\\
Supposing (A2) instead of (A1) does not bring anything to the expansion obtained in \eqref{suscepti5}, with the only difference that the functions $\mathfrak{d}_{j,l}(\cdot\,)$, $j=1,\ldots,\mathrm{Card}(\Theta)$ and $l=0,1,2,3$ (which are the same) are in this instance smooth w.r.t. the $\bold{k}$-variable. That is ensured by the analytic perturbation theory, see e.g. \cite[Sect. XII]{RS4} and also \cite[Thm 3.5]{N}. Nevertheless, considering the particular case of $\mathrm{Card}(\Theta)=2$ allows us to write down concise and explicit formulas for the functions appearing in an expansion of type \eqref{suscepti5}. Here is the counterpart of Proposition \ref{pro3.3} when $\mathrm{Card}(\Theta)=2$:

\begin{lema}
\label{complete2}
Let $\beta >0$ and $\rho_{0}>0$ be fixed. Let $\mu^{(0)} = \mu^{(0)}(\beta,\rho_{0},0) \in \mathbb{R}$ be the unique solution of the equation $\rho(\beta,\mathrm{e}^{\beta \mu},0) = \rho_{0}$. Then:
\begin{equation}
\label{exp2}
\mathcal{X}(\beta,\rho_{0},0) = - \frac{1}{4} \bigg(\frac{e}{c}\bigg)^{2} \frac{1}{2 \vert \Omega^{*}\vert} \frac{1}{\beta} \sum_{j=1}^{2} \int_{\Omega^{*}} \mathrm{d}\bold{k}\, \sum_{l=0}^{3} \frac{\partial^{l} \mathfrak{f}}{\partial \xi^{l}}(\beta,\mu^{(0)};E_{j}(\bold{k})) \hat{\mathfrak{d}}_{j,l}(\bold{k}),
\end{equation}
where $\forall\bold{k} \in \Omega^{*}$, the functions $\hat{\mathfrak{d}}_{j,l}(\cdot\,)$, $j=1,2$, $l=0,1,2,3$ are given by (below $m=1,2$, $m\neq j$):
\begin{gather}
\label{hatd3}
\hat{\mathfrak{d}}_{j,3}(\bold{k}):= \frac{1}{3!} \Big\{ \frac{\mathcal{C}_{j,j,m,j}(\bold{k})}{E_{m}(\bold{k}) - E_{j}(\bold{k})} + \mathcal{C}_{j,j,j}(\bold{k})\Big\},\\
\label{hatd2}
\hat{\mathfrak{d}}_{j,2}(\bold{k}):= - \frac{1}{2} \Big\{\frac{ \mathcal{C}_{j,j,m,m}(\bold{k}) + \mathcal{C}_{j,m,j,m}(\bold{k}) + \mathcal{C}_{j,m,m,j}(\bold{k})}{(E_{m}(\bold{k}) - E_{j}(\bold{k}))^{2}} +
\frac{\mathcal{C}_{j,m,j}(\bold{k}) + \mathcal{C}_{j,j,m}(\bold{k})}{E_{m}(\bold{k}) - E_{j}(\bold{k})} + \mathcal{C}_{j,j}(\bold{k})\Big\},\\
\label{hatd1-0}
\hat{\mathfrak{d}}_{j,1}(\bold{k}):= \sum_{n=0}^{2} \frac{\mathfrak{u}_{j,n}(\bold{k})}{(E_{m}(\bold{k}) - E_{j}(\bold{k}))^{n+1}},\qquad \hat{\mathfrak{d}}_{j,0}(\bold{k}):= \sum_{n=0}^{2} \frac{\mathfrak{v}_{j,n}(\bold{k})}{(E_{m}(\bold{k}) - E_{j}(\bold{k}))^{n+2}};
\end{gather}
and the functions $\mathfrak{u}_{j,n}(\cdot\,)$ and $\mathfrak{v}_{j,n}(\cdot\,)$, $j=1,2$, $n=0,1,2$ are given by (below $m=1,2$, $m\neq j$):
\begin{gather}
\label{u10}
\mathfrak{u}_{j,0}(\bold{k}) := \mathcal{C}_{j,m}(\bold{k}),\\
\label{u11}
\mathfrak{u}_{j,1}(\bold{k}) := \mathcal{C}_{j,m,m}(\bold{k}) + \mathcal{C}_{m,j,j}(\bold{k}) - \mathcal{C}_{j,m,j}(\bold{k}) - \mathcal{C}_{j,j,m}(\bold{k}),\\
\label{u12}
\mathfrak{u}_{j,2}(\bold{k}) := \mathcal{C}_{j,m,m,m}(\bold{k}) - \mathcal{C}_{m,j,j,j}(\bold{k}) - \mathcal{C}_{j,j,m,m}(\bold{k}) - \mathcal{C}_{j,m,j,m}(\bold{k}) - \mathcal{C}_{j,m,m,j}(\bold{k}),
\end{gather}
\begin{gather}
\label{v10}
\mathfrak{v}_{j,0}(\bold{k}) := 0,\\
\label{v11}
\mathfrak{v}_{j,1}(\bold{k}) := 2 \mathcal{C}_{j,m,m}(\bold{k}) + 2 \mathcal{C}_{m,j,j}(\bold{k}) - \mathcal{C}_{j,m,j}(\bold{k}) - \mathcal{C}_{j,j,m}(\bold{k}) - \mathcal{C}_{m,j,m}(\bold{k}) - \mathcal{C}_{m,m,j}(\bold{k}), \\
\label{v12}
\mathfrak{v}_{j,2}(\bold{k}) := 2\mathcal{C}_{j,m,m,m}(\bold{k}) - 2\mathcal{C}_{m,j,j,j}(\bold{k}).
\end{gather}
The functions $\mathcal{C}_{j_{1},j_{2},j_{3},j_{4}}(\cdot\,)$, $\mathcal{C}_{j_{1},j_{2},j_{3}}(\cdot\,)$ and $\mathcal{C}_{j_{1},j_{2}}(\cdot\,)$, $j_{1},j_{2},j_{3},j_{4}=1,2$ are defined in Lemma \ref{lemm1}.
\end{lema}

The above lemma is obtained from Lemma \ref{lemm1} by setting $\mathrm{Card}(\Theta)=2$, then by mimicking the proof of Lemmas \ref{lemm2}-\ref{lemm4}. Since this is simply calculations based on identities provided by the residue theorem, we do not give further details. Note that the above functions $\hat{\mathfrak{d}}_{j,l}(\cdot\,)$, with $j=1,2$, $l=0,1,2,3$ exactly have the same peculiarity than the one mentioned for the functions $\mathfrak{d}_{j,l}(\cdot\,)$ appearing in the expansion \eqref{suscepti5}, see  Proposition \ref{pro3.3}.\\

\indent The non-degeneracy assumption for the $E_{j}$'s, $j=1,2$ allows us to use the regular perturbation theory, which from the Hellman-Feynman formula (see e.g. \cite[Thm 4.1]{Hi}), provides the identities:
\begin{gather}
\label{derv1}
\frac{\partial E_{j}(\bold{k})}{\partial k_{\alpha}} = \hat{\pi}_{j,j}(\alpha;\bold{k}) \quad \alpha = 1,2,\\
\label{derv2}
\frac{\partial^{2} E_{j}(\bold{k})}{\partial k_{\alpha}^{2}} = \hat{\sigma}_{j,j}(\alpha,\alpha;\bold{k}) + 2 \sum_{m \neq j} \frac{\vert \hat{\pi}_{m,j}(\alpha;\bold{k})\vert^{2}}{E_{j}(\bold{k}) - E_{m}(\bold{k})} \quad \alpha =1,2,\\
\label{derv12}
\frac{\partial^{2} E_{j}(\bold{k})}{\partial k_{\alpha}\partial k_{\gamma}} = \hat{\sigma}_{j,j}(\alpha,\gamma;\bold{k}) + 2 \sum_{m\neq j} \frac{\Re\{ \hat{\pi}_{m,j}(\alpha;\bold{k}) \hat{\pi}_{j,m}(\gamma;\bold{k})\}}{E_{j}(\bold{k}) - E_{m}(\bold{k})} = \frac{\partial^{2} E_{j}(\bold{k})}{\partial k_{\gamma}\partial k_{\alpha}} \quad \alpha,\gamma =1,2.
\end{gather}
The derivatives of higher order can also be identified, but we will not use them in the following.\\
\indent The use of the regular perturbation theory plays a crucial role in the next result: it leads to the  identification of the interband contributions coming from the expansion \eqref{exp2}:

\begin{proposition}
\label{propo4}
Let $\beta >0$ and $\rho_{0}>0$ be fixed. Let $\mu^{(0)} = \mu^{(0)}(\beta,\rho_{0},0) \in \mathbb{R}$ be the unique solution of the equation $\rho(\beta,\mathrm{e}^{\beta \mu},0) = \rho_{0}$. Then \eqref{exp2} can be split into two contributions:
\begin{equation}
\label{exp4}
\mathcal{X}(\beta,\rho_{0},0) = \mathcal{X}_{\mathrm{P}}(\beta,\rho_{0},0) + \mathcal{X}_{\mathrm{Ib}}(\beta,\rho_{0},0),
\end{equation}
with, by using the shorthand notations $\frac{\partial^{l} \mathfrak{f}}{\partial \xi^{l}}(\beta,\mu^{(0)};\cdot\,) = \frac{\partial^{l} \mathfrak{f}}{\partial \xi^{l}}(\cdot\,)$, $l=0,1,2$:
\begin{gather}
\mathcal{X}_{\mathrm{P}}(\beta,\rho_{0},0):= - \frac{1}{48} \bigg(\frac{e}{c}\bigg)^{2} \frac{1}{\vert \Omega^{*}\vert} \frac{1}{\beta} \sum_{j=1}^{2} \int_{\Omega^{*}} \mathrm{d}\bold{k}\, \frac{\partial^{2} \mathfrak{f}}{\partial \xi^{2}}(E_{j}(\bold{k}))\bigg\{\frac{\partial^{2} E_{j}(\bold{k})}{\partial k_{1}^{2}} \frac{\partial^{2} E_{j}(\bold{k})}{\partial k_{2}^{2}} - \bigg(\frac{\partial^{2} E_{j}(\bold{k})}{\partial k_{1} \partial k_{2}}\bigg)^{2}\bigg\}, \nonumber \\
\label{ibc}
\mathcal{X}_{\mathrm{Ib}}(\beta,\rho_{0},0) :=
- \frac{1}{8} \bigg(\frac{e}{c}\bigg)^{2} \frac{1}{\vert \Omega^{*}\vert} \frac{1}{\beta} \sum_{j=1}^{2} \int_{\Omega^{*}} \mathrm{d}\bold{k}\, \bigg\{\frac{\partial^{2} \mathfrak{f}}{\partial \xi^{2}}(E_{j}(\bold{k})) \mathfrak{I}_{j}(\bold{k}) + \sum_{l=0}^{1} \frac{\partial^{l} \mathfrak{f}}{\partial \xi^{l}}(E_{j}(\bold{k})) \hat{\mathfrak{d}}_{j,l}(\bold{k})\bigg\},
\end{gather}
where $\forall\bold{k} \in \Omega^{*}$, the functions $\mathfrak{I}_{j}(\cdot\,)$, $j=1,2$ are given by (below $m=1,2$ with $m\neq j$):
\begin{equation}
\label{frakI}
\mathfrak{I}_{j}(\bold{k}) :=  \frac{1}{2} \Re \bigg\{-\frac{\mathcal{C}_{j,j,m,m}(\bold{k}) + \mathcal{C}_{j,m,j,m}(\bold{k}) + \mathcal{C}_{j,m,m,j}(\bold{k})}{(E_{j}(\bold{k}) - E_{m}(\bold{k}))^{2}} + i \frac{\Im \{\mathcal{C}_{j,m,j}(\bold{k})\}}{E_{j}(\bold{k}) - E_{m}(\bold{k})}\bigg\},
\end{equation}
and the functions $\hat{\mathfrak{d}}_{j,l}(\cdot\,)$, with $j=1,2$ and $l=0,1$ are defined in Lemma \ref{complete2}.
\end{proposition}

The proof of the above result can be found in the appendix of this section. The term $\mathcal{X}_{\mathrm{P}}(\beta,\rho_{0},0)$ is nothing but the so-called Peierls contribution, see e.g. \cite[Eq. (62c)]{Pei} and \cite[Eq. (4.10)]{BCS}. As for the term $\mathcal{X}_{\mathrm{Ib}}(\beta,\rho_{0},0)$, it stands for the interband (or band-to-band) contributions. 
As we will see below, only the functions multiplying $(\partial_{\xi}^{l} \mathfrak{f})(\beta,\mu^{(0)};\cdot\,)$ with $l=0,1$ (and therefore, only a great part of these interband contributions) will give rise to \eqref{xSC2}.\\

\indent We now can start the actual proof of Theorem \ref{thm2} $\mathrm{(iii)}$. By using the F-D distribution together with its partial derivatives, \eqref{exp2} reads as:
\begin{multline}
\label{derntfre}
\mathcal{X}(\beta,\rho_{0},0) = \frac{1}{8} \bigg(\frac{e}{c}\bigg)^{2} \frac{1}{\vert \Omega^{*}\vert} \sum_{j=1}^{2} \int_{\Omega^{*}} \mathrm{d}\bold{k}\, \bigg\{ \sum_{l=0}^{2} \frac{\partial^{l} \mathfrak{f}_{FD}}{\partial \xi^{l}}(\beta,\mu^{(0)};E_{j}(\bold{k})) \hat{\mathfrak{d}}_{j,l+1}(\bold{k}) + \\
- \frac{1}{\beta} \mathfrak{f}(\beta,\mu^{(0)};E_{j}(\bold{k})) \hat{\mathfrak{d}}_{j,0}(\bold{k})\bigg\}.
\end{multline}
Assume that the Fermi energy is in the middle of the gap separating the band $\mathcal{E}_{1}$ from $\mathcal{E}_{2}$. From \cite[Eqs. (4.2)-(4.3)]{BCS} together with \cite[Eq. (4.4)]{BCS} again, then all terms in \eqref{derntfre} containing derivatives of the Fermi-Dirac distribution will converge to zero in the limit $\beta \rightarrow \infty$. Hence, we get \eqref{xSC2}. Note that the expansion in \eqref{exp4} also leads to \eqref{xSC2} in the semiconducting situation.

\subsection{Appendix: proofs of the intermediate results.}

\noindent \textbf{Proof of Lemma \ref{lemm1}}. We start with the quantity $\mathcal{W}_{5}(\beta,\mu^{(0)},0)$.
Take a generical term of it:
\begin{multline*}
w_{5}(\beta,\mu^{(0)},0) := \frac{i}{2\pi} \int_{\Gamma} \mathrm{d}\xi\, \mathfrak{f}(\beta,\mu^{(0)};\xi) \int_{\Omega^{*}} \mathrm{d}\bold{k}\, \mathrm{Tr}_{l^{2}(\Theta)} (p_{5}(\bold{k};\xi)), \quad \textrm{with:}\\
p_{5}(\bold{k};\xi) := R(\bold{k};\xi) (\partial_{k_{\alpha}} H)(\bold{k}) R(\bold{k};\xi) (\partial_{k_{\gamma}} H)(\bold{k})R(\bold{k};\xi)(\partial_{k_{\gamma}} H)(\bold{k})R(\bold{k};\xi) (\partial_{k_{\alpha}} H)(\bold{k})R(\bold{k};\xi),
\end{multline*}
where $\alpha, \gamma = 1,2$. Without loss of generality, let us take for the moment $\gamma=\alpha$ (needed to simplify notations). By the mean of kernels, the trace of $p_{5}(\bold{k};\xi)$ can be written as:
\begin{equation}
\label{groform}
\sum_{\underline{\bold{x}} \in \Theta}\,\, \sum_{\underline{\bold{z}}_{1} \in \Theta} R(\underline{\bold{x}},\underline{\bold{z}}_{1};\bold{k},\xi) \bigg(\prod_{m=1}^{3} \sum_{\underline{\bold{z}}_{2m}, \underline{\bold{z}}_{2m+1} \in \Theta} (\partial_{k_{\alpha}} H)(\underline{\bold{z}}_{2m-1},\underline{\bold{z}}_{2m};\bold{k})
R(\underline{\bold{z}}_{2m},\underline{\bold{z}}_{2m+1};\bold{k},\xi)\bigg) R(\underline{\bold{z}}_{8},\underline{\bold{x}};\bold{k},\xi).
\end{equation}
Next use that the kernel of $R(\bold{k};\xi)$ can be expressed in terms of eigenfunctions of $H(\bold{k})$ as:
\begin{equation*}
\forall (\underline{\bold{x}},\underline{\bold{y}}) \in \Theta^{2},\quad R(\underline{\bold{x}},\underline{\bold{y}};\bold{k},\xi) = \sum_{j=1}^{\mathrm{Card}(\Theta)} \frac{u_{j}(\underline{\bold{x}};\bold{k}) \overline{u_{j}(\underline{\bold{y}};\bold{k})}}{E_{j}(\bold{k}) - \xi}\quad \bold{k} \in \Omega^{*}.
\end{equation*}
The replacement of each kernel of the resolvent in \eqref{groform} with the above expression yields:
\begin{multline*}
w_{5}(\beta,\mu^{(0)},0) = \sum_{j_{1},\ldots,j_{5} = 1}^{\mathrm{Card}(\Theta)} \frac{i}{2\pi} \int_{\Gamma} \mathrm{d}\xi\, \frac{\mathfrak{f}(\beta,\mu^{(0)};\xi)}{(E_{j_{1}}(\bold{k}) - \xi) \dotsb (E_{j_{5}}(\bold{k}) - \xi)} \int_{\Omega^{*}} \mathrm{d}\bold{k}\, \overbrace{\sum_{\underline{\bold{x}} \in \Theta} \overline{u_{j_{5}}(\underline{\bold{x}};\bold{k})} u_{j_{1}}(\underline{\bold{x}};\bold{k})}^{= \delta_{j_{1},j_{5}}} \times \\
\times \prod_{m=1}^{4} \sum_{\underline{\bold{z}}_{2m-1} \in \Theta} \overline{u_{j_{m}}(\underline{\bold{z}}_{2m-1};\bold{k})} \sum_{\underline{\bold{z}}_{2m} \in \Theta} (\partial_{k_{\alpha}} H)(\underline{\bold{z}}_{2m-1},\underline{\bold{z}}_{2m};\bold{k}) u_{j_{m+1}}(\underline{\bold{z}}_{2m};\bold{k}).
\end{multline*}
From the definition \eqref{defpi}, one therefore obtains:
\begin{equation}
\label{smw5}
w_{5}(\beta,\mu^{(0)},0) = \sum_{j_{1},\ldots,j_{4} = 1}^{\mathrm{Card}(\Theta)} \frac{i}{2\pi} \int_{\Gamma} \mathrm{d}\xi\, \frac{\mathfrak{f}(\beta,\mu^{(0)};\xi)}{(E_{j_{1}}(\bold{k}) - \xi)^{2}(E_{j_{2}}(\bold{k}) - \xi)\dotsb (E_{j_{4}}(\bold{k}) - \xi)}\int_{\Omega^{*}} \mathrm{d}\bold{k}\, \mathrm{c}_{j_{1},j_{2},j_{3},j_{4}}(\bold{k}),
\end{equation}
where $\mathrm{c}_{j_{1},j_{2},j_{3},j_{4}}(\bold{k}) := \hat{\pi}_{j_{1},j_{2}}(\alpha;\bold{k}) \hat{\pi}_{j_{2},j_{3}}(\gamma;\bold{k}) \hat{\pi}_{j_{3},j_{4}}(\gamma;\bold{k}) \hat{\pi}_{j_{4},j_{1}}(\alpha;\bold{k})$, $\bold{k} \in \Omega^{*}$. Let us now prove:
\begin{equation}
\label{kiesti}
\int_{\Gamma} \vert \mathrm{d}\xi\vert \, \bigg\vert \mathrm{c}_{j_{1},j_{2},j_{3},j_{4}}(\bold{k}) \frac{\mathfrak{f}(\beta,\mu^{(0)};\xi)}{(E_{j_{1}}(\bold{k}) - \xi)^{2}(E_{j_{2}}(\bold{k}) - \xi)\dotsb (E_{j_{4}}(\bold{k}) - \xi)} \bigg\vert \leq C,
\end{equation}
for some $\bold{k}$-independent constant $C>0$, what allows to change the order of integration in \eqref{smw5}. We only give the two main ingredients leading to \eqref{kiesti}. Let $\xi_{0} > 0$ so that $-\xi_{0} < \inf \sigma(H_{0})$ and large enough. On the one hand, from
\cite[Eq. (4.8)]{CN} there exists a $C = C_{\xi_{0}}(\beta,\mu^{(0)})>0$ s.t.:
\begin{equation*}
\int_{\Gamma} \vert \mathrm{d}\xi\vert\, \frac{\vert \mathfrak{f}(\beta,\mu^{(0)};\xi) \vert}{\vert E_{j_{1}}(\bold{k}) - \xi\vert^{2} \vert E_{j_{2}}(\bold{k}) - \xi\vert \dotsb \vert E_{j_{4}}(\bold{k}) - \xi\vert} \leq C (E_{j_{1}}(\bold{k}) + \xi_{0})^{-2}.
\end{equation*}
On the other hand, one has the following rough estimate. There exists another $C=C_{\xi_{0}}>0$ s.t.:
\begin{equation*}
\vert \hat{\pi}_{l,m}(\alpha;\bold{k}) \vert \leq C (E_{l}(\bold{k}) + \xi_{0}) \quad l,m =1,\ldots,\mathrm{Card}(\Theta),\, \alpha=1,2.
\end{equation*}
This follows from \cite[Eq. (4.6)]{CN} together with the fact that $(\partial_{k_{\alpha}} H)(\bold{k})$ is bounded from $l^{2}(\Theta)$ to $l^{\infty}(\Theta)$. To see that, use that $\Vert (\partial_{k_{\alpha}} H)(\bold{k})\Vert_{2,\infty}^{2} = \mathrm{ess} \sup_{\underline{\bold{x}} \in \Theta} \sum_{\underline{\bold{y}} \in \Theta} \vert (\partial_{k_{\alpha}} H)(\underline{\bold{x}},\underline{\bold{y}};\bold{k})\vert^{2}$ combined with \eqref{partialH} and the estimate \eqref{esH_0}. It remains to involve the continuity of the $E_{j}(\cdot\,)$'s to get \eqref{kiesti}.
To conclude the proof for the quantity $\mathcal{W}_{5}(\beta,\mu^{(0)},0)$, it is sufficient to use that it is made up of a linear combination of $w_{5}(\beta,\mu^{(0)},0)$-like terms. The quantities $\mathcal{W}_{l}(\beta,\mu^{(0)},0)$, with $l=3,4$ can be treated with similar arguments.\qed \\

\noindent\textbf{Proof of Lemma \ref{lemm2}.} Denote the integrand appearing in \eqref{trP5} by:
\begin{equation*}
\mathfrak{g}_{j_{1},j_{2},j_{3},j_{4}}(\beta,\mu^{(0)};\xi) := \frac{\mathfrak{f}(\beta,\mu^{(0)};\xi)}{(E_{j_{1}}(\bold{k}) - \xi)^{2} (E_{j_{2}}(\bold{k}) - \xi) (E_{j_{3}}(\bold{k}) - \xi) (E_{j_{4}}(\bold{k}) - \xi)} \quad j_{1},j_{2},j_{3},j_{4} \in \mathbb{N}^{*}.
\end{equation*}
At first sight, $\mathfrak{g}_{j_{1},j_{2},j_{3},j_{4}}(\beta,\mu^{(0)};\cdot\,)$ can have poles from the first order up to at most fifth order (in the case when $j_{1}=j_{2}=j_{3}=j_{4}$). Therefore we expect the integral w.r.t. $\xi$ of $\mathfrak{g}_{j_{1},j_{2},j_{3},j_{4}}(\beta,\mu^{(0)};\cdot\,)$ to make appear partial derivatives of $\mathfrak{f}(\beta,\mu^{(0)};\cdot\,)$ with order at most equal to four. However the factor multiplying $(\partial_{\xi}^{4}\mathfrak{f})(\beta,\mu^{(0)};\cdot\,)$ is identically zero. Indeed, in view of the function $\mathcal{C}_{j_{1},j_{2},j_{3},j_{4}}(\cdot\,)$ defined in \eqref{C4}, it is identically zero for the combinations of subscripts:
\begin{equation}
\label{combi}
j_{1}=j_{2}=j_{3}=j_{4};\quad j_{1}=j_{2}=j_{3}\neq j_{4};\quad j_{1}=j_{3}=j_{4}\neq j_{2}.
\end{equation}
Ergo the expansion of \eqref{trP5} consists of partial derivatives of $\mathfrak{f}(\beta,\mu^{(0)};\cdot\,)$ of order at most equal to three. Then by virtue of \eqref{combi}, the quadruple summation in the r.h.s. of \eqref{trP5} is reduced to:
\begin{multline*}
\mathcal{W}_{5}(\beta,\mu^{(0)},0) = - \sum_{j_{1}=1}^{\mathrm{Card}(\Theta)} \sum_{\substack{j_{3}=1 \\ j_{3} \neq j_{1}}}^{\mathrm{Card}(\Theta)} \int_{\Omega^{*}} \mathrm{d}\bold{k}\, \mathcal{C}_{j_{1},j_{1},j_{3},j_{1}}(\bold{k}) \frac{1}{2i\pi} \int_{\Gamma} \mathrm{d}\xi\, \frac{\mathfrak{f}(\beta,\mu^{(0)};\xi)}{(E_{j_{1}}(\bold{k}) - \xi)^{4} (E_{j_{3}}(\bold{k}) - \xi)} + \\
- \sum_{\underbrace{j_{1},\ldots,j_{4}=1}_{\substack{\textrm{at most 3 equal} \\ \textrm{subscripts}}}}^{\mathrm{Card}(\Theta)} \int_{\Omega^{*}} \mathrm{d}\bold{k}\, \mathcal{C}_{j_{1},j_{2},j_{3},j_{4}}(\bold{k}) \frac{1}{2i\pi} \int_{\Gamma} \mathrm{d}\xi\, \frac{\mathfrak{f}(\beta,\mu^{(0)};\xi)}{(E_{j_{1}}(\bold{k}) - \xi)^{2}(E_{j_{2}}(\bold{k}) - \xi)(E_{j_{3}}(\bold{k}) - \xi)(E_{j_{4}}(\bold{k})-\xi)}.
\end{multline*}
By applying the residue theorem in the first term of the above r.h.s., one has:
\begin{multline*}
\sum_{\substack{j_{3}=1 \\ j_{3} \neq j_{1}}}^{\mathrm{Card}(\Theta)} \int_{\Omega^{*}} \mathrm{d}\bold{k}\, \mathcal{C}_{j_{1},j_{1},j_{3},j_{1}}(\bold{k}) \frac{1}{2i\pi} \int_{\Gamma} \mathrm{d}\xi\, \frac{\mathfrak{f}(\beta,\mu^{(0)};\xi)}{(E_{j_{1}}(\bold{k}) - \xi)^{4} (E_{j_{3}}(\bold{k}) - \xi)} =   \sum_{\substack{j_{3}=1 \\ j_{3} \neq j_{1}}}^{\mathrm{Card}(\Theta)} \int_{\Omega^{*}} \mathrm{d}\bold{k}\, \mathcal{C}_{j_{1},j_{1},j_{3},j_{1}}(\bold{k}) \times \\
\times \bigg\{ \frac{1}{3!} \frac{1}{E_{j_{3}}(\bold{k}) - E_{j_{1}}(\bold{k})} \frac{\partial^{3} \mathfrak{f}}{\partial \xi^{3}}(\beta,\mu^{(0)};E_{j_{1}}(\bold{k}))
+ \textrm{others terms involving $\displaystyle{\frac{\partial^{l} \mathfrak{f}}{\partial \xi^{l}}(\beta,\mu^{(0)};\cdot\,)}$ with $l \leq 2$}\bigg\}.
\end{multline*}
The function $\mathcal{C}_{j_{1},j_{1},j_{3},j_{1}}(\cdot\,)$ appearing in front of $(\partial_{\xi}^{3} \mathfrak{f})(\beta,\mu^{(0)};E_{j_{1}}(\bold{k}))$ corresponds to $3! \mathfrak{a}_{j_{1},3}(\cdot\,)$. The remainder of the proof only is a plain computation using the residue theorem.\qed \\

\noindent \textbf{Proof of Lemma \ref{lemm3}.} Denote the integrand appearing in \eqref{trP4} by:
\begin{equation*}
\mathfrak{h}_{j_{1},j_{2},j_{3}}(\beta,\mu^{(0)};\xi) := \frac{\mathfrak{f}(\beta,\mu^{(0)};\xi)}{(E_{j_{1}}(\bold{k}) - \xi)^{2} (E_{j_{2}}(\bold{k}) - \xi) (E_{j_{3}}(\bold{k}) - \xi)} \quad j_{1},j_{2},j_{3} \in \mathbb{N}^{*}.
\end{equation*}
Note that $\mathfrak{h}_{j_{1},j_{2},j_{3}}(\beta,\mu^{(0)};\cdot\,)$ can have poles from the first order up to at most fourth order (in the case when $j_{1}=j_{2}=j_{3}$). Therefore we expect the integral w.r.t. $\xi$ of $\mathfrak{h}_{j_{1},j_{2},j_{3}}(\beta,\mu^{(0)};\cdot\,)$ to make appear partial derivatives of $\mathfrak{f}(\beta,\mu^{(0)};\cdot\,)$ with order at most equal to three.\\
The triple summation in the r.h.s. of \eqref{trP4} can be rewritten:
\begin{multline*}
\mathcal{W}_{4}(\beta,\mu^{(0)},0) = - \sum_{j_{1}=1}^{\mathrm{Card}(\Theta)} \int_{\Omega^{*}} \mathrm{d}\bold{k}\, \mathcal{C}_{j_{1},j_{1},j_{1}}(\bold{k}) \frac{1}{2i\pi} \int_{\Gamma} \mathrm{d}\xi\, \frac{\mathfrak{f}(\beta,\mu^{(0)};\xi)}{(E_{j_{1}}(\bold{k}) - \xi)^{4}} + \\
- \sum_{\underbrace{j_{1},j_{2},j_{3}=1}_{\substack{\textrm{at most 2 equal} \\ \textrm{subscripts}}}}^{\mathrm{Card}(\Theta)} \int_{\Omega^{*}} \mathrm{d}\bold{k}\, \mathcal{C}_{j_{1},j_{2},j_{3}}(\bold{k}) \frac{1}{2i\pi} \int_{\Gamma} \mathrm{d}\xi\, \frac{\mathfrak{f}(\beta,\mu^{(0)};\xi)}{(E_{j_{1}}(\bold{k}) - \xi)^{2}(E_{j_{2}}(\bold{k}) - \xi)(E_{j_{3}}(\bold{k}) - \xi)}.
\end{multline*}
The application of the residue theorem in the first term of the above r.h.s. yields:
\begin{equation*}
\frac{1}{2i\pi} \int_{\Gamma} \mathrm{d}\xi\, \frac{\mathfrak{f}(\beta,\mu^{(0)};\xi)}{(E_{j_{1}}(\bold{k}) - \xi)^{4}} = \frac{1}{3!} \frac{\partial^{3} \mathfrak{f}}{\partial \xi^{3}}(\beta,\mu^{(0)};E_{j_{1}}(\bold{k})).
\end{equation*}
This is only the one term which provides a third-order partial derivative of $\mathfrak{f}(\beta,\mu^{(0)};\cdot\,)$. \qed \\

\noindent \textbf{Proof of Lemma \ref{lemm4}.} Denote the integrand appearing in \eqref{trP3} by:
\begin{equation*}
\mathfrak{i}_{j_{1},j_{2}}(\beta,\mu^{(0)};\xi) := \frac{\mathfrak{f}(\beta,\mu^{(0)};\xi)}{(E_{j_{1}}(\bold{k}) - \xi)^{2} (E_{j_{2}}(\bold{k}) - \xi)} \quad j_{1},j_{2} \in \mathbb{N}^{*}.
\end{equation*}
Note that $\mathfrak{i}_{j_{1},j_{2}}(\beta,\mu^{(0)};\cdot\,)$ can have poles from the first order up to at most third order (in the case when $j_{1}=j_{2}$). Therefore the integral w.r.t. $\xi$ of $\mathfrak{i}_{j_{1},j_{2}}(\beta,\mu^{(0)};\cdot\,)$ will make appear partial derivatives of $\mathfrak{f}(\beta,\mu^{(0)};\cdot\,)$ with order at most equal to two. We do not give further details. \qed \\

\noindent \textbf{Proof of Proposition \ref{propo4}}.
Consider the quantity $\hat{\mathfrak{d}}_{j,3}(\bold{k})$, $j=1,2$ defined in \eqref{hatd3}. Let $m=1,2$ with $m \neq j$. From \eqref{C4} and \eqref{C3} together with the identities \eqref{derv1}-\eqref{derv12} we have:
\begin{multline*}
\frac{\mathcal{C}_{j,j,m,j}(\bold{k})}{E_{m}(\bold{k}) - E_{j}(\bold{k})} = \bigg(\frac{\partial E_{j}(\bold{k})}{\partial k_{1}}\bigg)^{2}\frac{1}{2} \bigg[\hat{\sigma}_{j,j}(2,2;\bold{k}) - \frac{\partial^{2} E_{j}(\bold{k})}{\partial k_{2}^{2}}\bigg] + \\
+ \bigg(\frac{\partial E_{j}(\bold{k})}{\partial k_{2}}\bigg)^{2}\frac{1}{2} \bigg[\hat{\sigma}_{j,j}(1,1;\bold{k}) - \frac{\partial^{2} E_{j}(\bold{k})}{\partial k_{1}^{2}}\bigg] - \bigg(\frac{\partial E_{j}(\bold{k})}{\partial k_{1}}\bigg)\bigg(\frac{\partial E_{j}(\bold{k})}{\partial k_{2}}\bigg) \bigg[\hat{\sigma}_{j,j}(1,2;\bold{k}) - \frac{\partial^{2} E_{j}(\bold{k})}{\partial k_{1} \partial k_{2}}\bigg],
\end{multline*}
\begin{equation*}
\mathcal{C}_{j,j,j}(\bold{k}) = \hat{\sigma}_{j,j}(1,1;\bold{k}) \bigg(\frac{\partial E_{j}(\bold{k})}{\partial k_{2}}\bigg)^{2} + \hat{\sigma}_{j,j}(2,2;\bold{k}) \bigg(\frac{\partial E_{j}(\bold{k})}{\partial k_{1}}\bigg)^{2} - 2 \hat{\sigma}_{j,j}(1,2;\bold{k}) \bigg(\frac{\partial E_{j}(\bold{k})}{\partial k_{1}}\bigg)\bigg(\frac{\partial E_{j}(\bold{k})}{\partial k_{2}}\bigg).
\end{equation*}
Thus the quantity $\hat{\mathfrak{d}}_{j,3}(\bold{k})$ can be rewritten
$\hat{\mathfrak{d}}_{j,3}(\bold{k}) = \mathfrak{m}_{j,3}(\bold{k}) + \mathfrak{n}_{j,3}(\bold{k})$, $\bold{k} \in \Omega^{*}$
where respectively:
\begin{gather*}
\mathfrak{m}_{j,3}(\bold{k}) := \frac{-1}{12}\bigg\{\bigg(\frac{\partial E_{j}(\bold{k})}{\partial k_{1}}\bigg)^{2} \frac{\partial^{2} E_{j}(\bold{k})}{\partial k_{2}^{2}} + \bigg(\frac{\partial E_{j}(\bold{k})}{\partial k_{2}}\bigg)^{2} \frac{\partial^{2} E_{j}(\bold{k})}{\partial k_{1}^{2}} - 2 \bigg(\frac{\partial E_{j}(\bold{k})}{\partial k_{1}}\bigg)\bigg(\frac{\partial E_{j}(\bold{k})}{\partial k_{2}}\bigg) \frac{\partial^{2} E_{j}(\bold{k})}{\partial k_{1}\partial k_{2}}\bigg\}, \\
\mathfrak{n}_{j,3}(\bold{k}) := \frac{1}{4}\bigg\{\bigg(\frac{\partial E_{j}(\bold{k})}{\partial k_{1}}\bigg)^{2} \hat{\sigma}_{j,j}(2,2;\bold{k}) + \bigg(\frac{\partial E_{j}(\bold{k})}{\partial k_{2}}\bigg)^{2} \hat{\sigma}_{j,j}(1,1;\bold{k}) - 2\bigg(\frac{\partial E_{j}(\bold{k})}{\partial k_{1}}\bigg)\bigg(\frac{\partial E_{j}(\bold{k})}{\partial k_{2}}\bigg) \hat{\sigma}_{j,j}(1,2;\bold{k}) \bigg\}.
\end{gather*}
Hereafter we use the shorthand notations $\mathfrak{f}(\cdot\,) = \mathfrak{f}(\beta,\mu^{(0)};\cdot\,)$. Let us now prove the following:
\begin{equation}
\label{rfity}
\int_{\Omega^{*}} \mathrm{d}\bold{k}\, \frac{\partial^{3}\mathfrak{f}}{\partial \xi^{3}}(E_{j}(\bold{k})) \mathfrak{m}_{j,3}(\bold{k}) =
\frac{1}{6} \int_{\Omega^{*}} \mathrm{d}\bold{k}\, \frac{\partial^{2}\mathfrak{f}}{\partial \xi^{2}}(E_{j}(\bold{k})) \bigg\{ \frac{\partial^{2} E_{j}(\bold{k})}{\partial k_{1}^{2}} \frac{\partial^{2} E_{j}(\bold{k})}{\partial k_{2}^{2}} - \bigg(\frac{\partial^{2} E_{j}(\bold{k})}{\partial k_{1} \partial k_{2}}\bigg)^{2}\bigg\}.
\end{equation}
From \cite[Eq. (4.15)]{BCS} and after rearranging of terms, one has:
\begin{multline*}
\int_{\Omega^{*}} \mathrm{d}\bold{k}\, \frac{\partial^{3} \mathfrak{f}}{\partial \xi^{3}}(E_{j}(\bold{k})) \mathfrak{m}_{j,3}(\bold{k}) = \frac{1}{6} \int_{\Omega^{*}} \mathrm{d}\bold{k}\, \frac{\partial^{3} \mathfrak{f}}{\partial \xi^{3}}(E_{j}(\bold{k})) \bigg\{\bigg(\frac{\partial E_{j}(\bold{k})}{\partial k_{1}}\bigg)\bigg(\frac{\partial E_{j}(\bold{k})}{\partial k_{2}}\bigg) \frac{\partial^{2} E_{j}(\bold{k})}{\partial k_{1} \partial k_{2}}\bigg\} + \\
+ \frac{1}{12} \int_{\Omega^{*}} \mathrm{d}\bold{k}\, \frac{\partial^{2} \mathfrak{f}}{\partial \xi^{2}}(E_{j}(\bold{k})) \bigg\{2 \frac{\partial^{2} E_{j}(\bold{k})}{\partial k_{1}^{2}} \frac{\partial^{2} E_{j}(\bold{k})}{\partial k_{2}^{2}} + \frac{\partial E_{j}(\bold{k})}{\partial k_{1}} \frac{\partial}{\partial k_{2}} \frac{\partial^{2} E_{j}(\bold{k})}{\partial k_{1} \partial k_{2}} + \frac{\partial E_{j}(\bold{k})}{\partial k_{2}} \frac{\partial}{\partial k_{1}} \frac{\partial^{2} E_{j}(\bold{k})}{\partial k_{2} \partial k_{1}} \bigg\}.
\end{multline*}
The above identity corresponds to \cite[Eq. (4.16)]{BCS} added with \cite[Eq. (4.13)]{BCS} by setting $N=j$. By following the same method as the one concluding the proof of \cite[Prop. 4.2]{BCS}, one obtains \eqref{rfity}.
Next, let us consider \eqref{hatd2}. In view of \eqref{frakI} what we still have to do is to prove that:
\begin{multline}
\label{rfgth}
\int_{\Omega^{*}} \mathrm{d}\bold{k}\, \frac{\partial^{3} \mathfrak{f}}{\partial \xi^{3}}(E_{j}(\bold{k})) \mathfrak{n}_{j,3}(\bold{k}) - \frac{1}{2} \int_{\Omega^{*}} \mathrm{d}\bold{k}\, \frac{\partial^{2} \mathfrak{f}}{\partial \xi^{2}}(E_{j}(\bold{k})) \bigg\{ \frac{\mathcal{C}_{j,m,j}(\bold{k}) + \mathcal{C}_{j,j,m}(\bold{k})}{E_{m}(\bold{k}) - E_{j}(\bold{k})} + \mathcal{C}_{j,j}(\bold{k})\bigg\} = \\ \frac{i}{2} \int_{\Omega^{*}} \mathrm{d}\bold{k}\, \frac{\partial^{2} \mathfrak{f}}{\partial \xi^{2}}(E_{j}(\bold{k}))
\frac{ \Im\{ \mathcal{C}_{j,m,j}(\bold{k})\}}{E_{j}(\bold{k}) - E_{m}(\bold{k})}.
\end{multline}
On the one hand, from the definition \eqref{C3} together with the identities \eqref{derv1}-\eqref{derv12}:
\begin{multline*}
\frac{\mathcal{C}_{j,j,m}(\bold{k})}{E_{m}(\bold{k}) - E_{j}(\bold{k})} = \hat{\sigma}_{j,j}(1,1;\bold{k})\bigg[\frac{1}{2} \hat{\sigma}_{j,j}(2,2;\bold{k}) - \frac{1}{2} \frac{\partial^{2} E_{j}(\bold{k})}{\partial k_{2}^{2}}\bigg] + \\
+ \hat{\sigma}_{j,j}(2,2;\bold{k})\bigg[\frac{1}{2} \hat{\sigma}_{j,j}(1,1;\bold{k}) - \frac{1}{2} \frac{\partial^{2} E_{j}(\bold{k})}{\partial k_{1}^{2}}\bigg]
- \hat{\sigma}_{j,j}(1,2;\bold{k})\bigg[\hat{\sigma}_{j,j}(1,2;\bold{k}) - \frac{\partial^{2} E_{j}(\bold{k})}{\partial k_{1} \partial k_{2}}\bigg].
\end{multline*}
Then from the definition \eqref{C2} we get after rearranging of terms:
\begin{multline}
\label{gthf1}
- \frac{1}{2} \int_{\Omega^{*}} \mathrm{d}\bold{k}\, \frac{\partial^{2} \mathfrak{f}}{\partial \xi^{2}}(E_{j}(\bold{k}))\bigg\{ \frac{\mathcal{C}_{j,m,j}(\bold{k}) + \mathcal{C}_{j,j,m}(\bold{k})}{E_{m}(\bold{k}) - E_{j}(\bold{k})} + \mathcal{C}_{j,j}(\bold{k})\bigg\} =  \frac{1}{4} \int_{\Omega^{*}} \mathrm{d}\bold{k}\, \frac{\partial^{2} \mathfrak{f}}{\partial \xi^{2}}(E_{j}(\bold{k})) \times \\
\times \bigg\{ \frac{\partial^{2} E_{j}(\bold{k})}{\partial k_{1}^{2}} \hat{\sigma}_{j,j}(2,2;\bold{k})
+ \frac{\partial^{2} E_{j}(\bold{k})}{\partial k_{2}^{2}} \hat{\sigma}_{j,j}(1,1;\bold{k}) - 2 \frac{\partial^{2} E_{j}(\bold{k})}{\partial k_{1} \partial k_{2}} \hat{\sigma}_{j,j}(1,2;\bold{k}) -  2 \frac{\mathcal{C}_{j,m,j}(\bold{k})}{E_{m}(\bold{k}) - E_{j}(\bold{k})}\bigg\}.
\end{multline}
On the other hand, the definition of $\mathfrak{n}_{j,3}(\bold{k})$ together with the use of
\cite[Eq. (4.15)]{BCS} leads to:
\begin{multline}
\label{gthf2}
\int_{\Omega^{*}} \mathrm{d}\bold{k}\, \frac{\partial^{3} \mathfrak{f}}{\partial \xi^{3}}(E_{j}(\bold{k})) \mathfrak{n}_{j,3}(\bold{k}) = - \frac{1}{4} \int_{\Omega^{*}} \mathrm{d}\bold{k}\, \frac{\partial^{2} \mathfrak{f}}{\partial \xi^{2}}(E_{j}(\bold{k})) \bigg\{
\frac{\partial}{\partial k_{1}} \bigg(\frac{\partial E_{j}(\bold{k})}{\partial k_{1}} \hat{\sigma}_{j,j}(2,2;\bold{k})\bigg) + \\
+ \frac{\partial}{\partial k_{2}} \bigg( \frac{\partial E_{j}(\bold{k})}{\partial k_{2}} \hat{\sigma}_{j,j}(1,1;\bold{k})\bigg) - \frac{\partial}{\partial k_{1}} \bigg(\frac{\partial E_{j}(\bold{k})}{\partial k_{2}} \hat{\sigma}_{j,j}(1,2;\bold{k})\bigg) - \frac{\partial}{\partial k_{2}} \bigg(\frac{\partial E_{j}(\bold{k})}{\partial k_{1}} \hat{\sigma}_{j,j}(1,2;\bold{k})\bigg)\bigg\}.
\end{multline}
By adding \eqref{gthf1} with \eqref{gthf2} and using the definition of $\mathcal{C}_{j,m,j}(\bold{k})$, the l.h.s. of \eqref{rfgth} is equal to:
\begin{multline*}
\frac{1}{2} \int_{\Omega^{*}} \mathrm{d}\bold{k}\, \frac{\partial^{2} \mathfrak{f}}{\partial \xi^{2}}(E_{j}(\bold{k}))  \bigg\{ \frac{\partial E_{j}(\bold{k})}{\partial k_{1}} \bigg(\frac{\hat{\sigma}_{j,m}(2,2;\bold{k})\hat{\pi}_{m,j}(1;\bold{k})}{E_{j}(\bold{k}) - E_{m}(\bold{k})} - \frac{1}{2} \frac{\partial}{\partial k_{1}} \hat{\sigma}_{j,j}(2,2;\bold{k}) + \\
- \frac{\hat{\sigma}_{j,m}(1,2;\bold{k})\hat{\pi}_{m,j}(2;\bold{k})}{E_{j}(\bold{k}) - E_{m}(\bold{k})}  + \frac{1}{2} \frac{\partial}{\partial k_{2}} \hat{\sigma}_{j,j}(1,2;\bold{k})\bigg)
+ \frac{\partial E_{j}(\bold{k})}{\partial k_{2}} \bigg(\frac{\hat{\sigma}_{j,m}(1,1;\bold{k})\hat{\pi}_{m,j}(2;\bold{k})}{E_{j}(\bold{k}) - E_{m}(\bold{k})}
+ \\ - \frac{1}{2} \frac{\partial}{\partial k_{2}} \hat{\sigma}_{j,j}(1,1;\bold{k}) - \frac{\hat{\sigma}_{j,m}(1,2;\bold{k})\hat{\pi}_{m,j}(1;\bold{k})}{E_{j}(\bold{k}) - E_{m}(\bold{k})}  + \frac{1}{2} \frac{\partial}{\partial k_{1}} \hat{\sigma}_{j,j}(1,2;\bold{k})\bigg)\bigg\}.
\end{multline*}
To get \eqref{rfgth}, it remains to use this identity (derived from the regular perturbation theory):
\begin{equation*}
\frac{\partial}{\partial k_{\delta}} \hat{\sigma}_{j,j}(\alpha,\gamma;\bold{k}) = \hat{\tau}_{j,j}(\delta,\alpha,\gamma;\bold{k}) + \frac{2 \Re \{\hat{\sigma}_{j,m}(\alpha,\gamma;\bold{k}) \hat{\pi}_{m,j}(\delta;\bold{k})\}}{E_{j}(\bold{k}) - E_{m}(\bold{k})}\quad \bold{k} \in \Omega^{*},\,\alpha,\gamma,\delta = 1,2,
\end{equation*}
with $\hat{\tau}_{j,j}(\alpha,\alpha,\gamma;\bold{k}) := \langle u_{j}(\cdot\,;\bold{k}),(\partial_{k_{\alpha}} \partial_{k_{\alpha}} \partial_{k_{\gamma}} H)(\bold{k}) u_{j}(\cdot\,;\bold{k})\rangle = \hat{\tau}_{j,j}(\alpha,\gamma,\alpha;\bold{k}) = \hat{\tau}_{j,j}(\gamma,\alpha,\alpha;\bold{k})$.\qed

\section{Acknowledgments.}

B.S. warmly thanks Horia D. Cornean for many stimulating and fruitful discussions. A part of this work was done while the author was visiting the Department of Mathematical Sciences of Aalborg University. B.S. is thankful for the hospitality and financial support from the Department. As well, B.S. is indebted to Mathieu Beau for making many references available to the author.

\end{document}